\documentclass[letter,longauth]{aa}

\usepackage{natbib}
\usepackage{graphicx}
\usepackage{txfonts}
\usepackage[bookmarks=false, colorlinks=true, citecolor=blue, linkcolor=blue]{hyperref}
\usepackage{subfig}
\usepackage{blindtext}

\newcommand{\JWST}{JWST}
\newcommand{\HST}{HST}
\def\micron{\hbox{\,$\mu$m}}

\newcommand{\Msun}{\hbox{$M_{\rm \odot}$}}

\newcommand\nodata{ ~$\cdots$~ }

\newcommand{\Neii}{\hbox{[\ion{Ne}{ii}]12.81\micron}}
\newcommand{\Neiii}{\hbox{[\ion{Ne}{iii}]15.56\micron}}
\newcommand{\Neva}{\hbox{[\ion{Ne}{v}]14.32\micron}}
\newcommand{\Nevb}{\hbox{[\ion{Ne}{v}]24.32\micron}}
\newcommand{\Nevi}{\hbox{[\ion{Ne}{vi}]7.65\micron}}
\newcommand{\Arii}{\hbox{[\ion{Ar}{ii}]6.99\micron}}
\newcommand{\Mgv}{\hbox{[\ion{Mg}{v}]5.61\micron}}
\newcommand{\Feii}{\hbox{[\ion{Fe}{ii}]5.34\micron}}

\bibpunct{(}{)}{;}{a}{}{,}

\DeclareCaptionListOfFormat{subreformat}{#1#2}
\titlerunning{Low-power jet--ISM interaction revealed by \JWST\slash MIRI MRS}
\authorrunning{Pereira-Santaella et al.}

\begin{document}

\title{Low-power jet--interstellar medium interaction in NGC~7319 revealed by JWST\slash MIRI MRS}

\author{M.~Pereira-Santaella\inst{\ref{inst1},\ref{inst2}} \and J.~\'Alvarez-M\'arquez\inst{\ref{inst1}} 
\and I.~Garc\'ia-Bernete\inst{\ref{inst3}} 
\and A.~Labiano\inst{\ref{inst4},\ref{inst5}}
\and L.~Colina\inst{\ref{inst1}}
\and A.~Alonso-Herrero\inst{\ref{inst5}}
\and E.~Bellocchi\inst{\ref{inst6},\ref{inst7}}
\and S.~Garc\'ia-Burillo\inst{\ref{inst2}}
\and S.~F. H\"onig\inst{\ref{inst8}}
\and C.~Ramos Almeida\inst{\ref{inst9},\ref{inst10}}
\and D.~Rosario\inst{\ref{inst11}}
}
\institute{Centro de Astrobiolog\'ia (CSIC-INTA), Ctra de Torrej\'on a Ajalvir, km 4, 28850, Torrej\'on de Ardoz, Madrid, Spain\label{inst1}
\and 
Observatorio Astron\'omico Nacional (OAN-IGN)-Observatorio de Madrid, Alfonso XII, 3, 28014, Madrid, Spain\\ \email{miguel.pereira@oan.es}\label{inst2}
\and 
Department of Physics, University of Oxford, Keble Road, Oxford OX1 3RH, UK\label{inst3}
\and
Telespazio UK for the European Space Agency, ESAC, Camino Bajo del Castillo s/n, 28692 Villanueva de la Ca\~nada, Spain\label{inst4}
\and
Centro de Astrobiolog\'{\i}a (CSIC-INTA), ESAC  Campus, E-28692 Villanueva de la Ca\~nada, Madrid, Spain\label{inst5}
\and
Departamento de F\'isica de la Tierra y Astrof\'isica, Fac. de CC F\'isicas, Universidad Complutense de Madrid, 28040, Madrid, Spain\label{inst6}
\and
Instituto de F\'isica de Part\'iculas y del Cosmos IPARCOS, Fac. CC F\'isicas, Universidad Complutense de Madrid, 28040 Madrid, Spain\label{inst7}
\and
Department of Physics \& Astronomy, University of Southampton, Hampshire, SO17 1BJ, Southampton, UK\label{inst8}
\and
Instituto de Astrof\'isica de Canarias, Calle V\'ia Láctea, s/n, 38205 La Laguna, Tenerife, Spain\label{inst9}
\and
Departamento de Astrof\'isica, Universidad de La Laguna, 38206, La Laguna, Tenerife, Spain\label{inst10}
\and
School of Mathematics, Statistics and Physics, Newcastle University, NE1 7RU, Newcastle upon Tyne, UK\label{inst11}
}

\abstract{
We present JWST\slash MIRI MRS spectroscopy of NGC~7319, the largest galaxy in the Stephan's Quintet, observed as part of the Early Release Observations (ERO). NGC~7319 hosts a type 2 active galactic nucleus (AGN) and a low-power radio jet ($L_{\rm 1.4\,GHz}$=3.3$\times$10$^{22}$\,W\,Hz$^{-1}$) with two asymmetric radio hotspots at 430\,pc (N2) and 1.5\,kpc (S2) projected distances from the unresolved radio core.
The MRS data suggest that the molecular material in the disk of the galaxy decelerates the jet and causes this length asymmetry.
We find enhanced emission from warm and hot H$_2$ ($T_{\rm w}$=330$\pm$40\,K, $T_{\rm h}$=900$\pm$60\,K) and ionized gas at the intersection between the jet axis and dust lanes in the disk. This emission is coincident with the radio hotspot N2, the hotspot closer to the core, suggesting that the jet--interstellar medium (ISM) interaction decelerates the jet. Conversely, the mid-infrared emission at the more distant hotspot is fainter, more highly ionized, and with lower H$_2$ excitation, suggesting a more diffuse atomic environment where the jet can progress to farther distances.
At the N2 radio hotspot, the ionized gas mass ($M_{\rm ion}$=(2.4--12)$\times$10$^5$\,\Msun) is comparable to that of the warm H$_2$, but the former is more turbulent ($\sigma_{\rm ion}\sim300$ vs. $\sigma_{\rm H_2}\sim150$\,km\,s$^{-1}$), so the mechanical energy of the ionized gas is $\sim$1.3--10 times higher. From these estimates, we find that only {$<$1\%} of the jet energy remains as mechanical energy in these two ISM phases at N2.
We also find extended ($r>$0.3--1.5\,kpc) high-ionization emission ([\ion{Mg}{v}], [\ion{Ne}{vi}], and [\ion{Ne}{v}]) close to the radio hotspots.

This initial analysis of NGC~7319 shows the potential of MIRI\slash MRS to investigate the AGN feedback mechanisms due to radio jets and their radiation field in the, often heavily dust-enshrouded, central regions of galaxies. Understanding these mechanisms is an essential ingredient in the development of cosmological simulations of galaxy evolution.
}

\maketitle

\begin{figure*}
\centering
\includegraphics[width=0.75\textwidth]{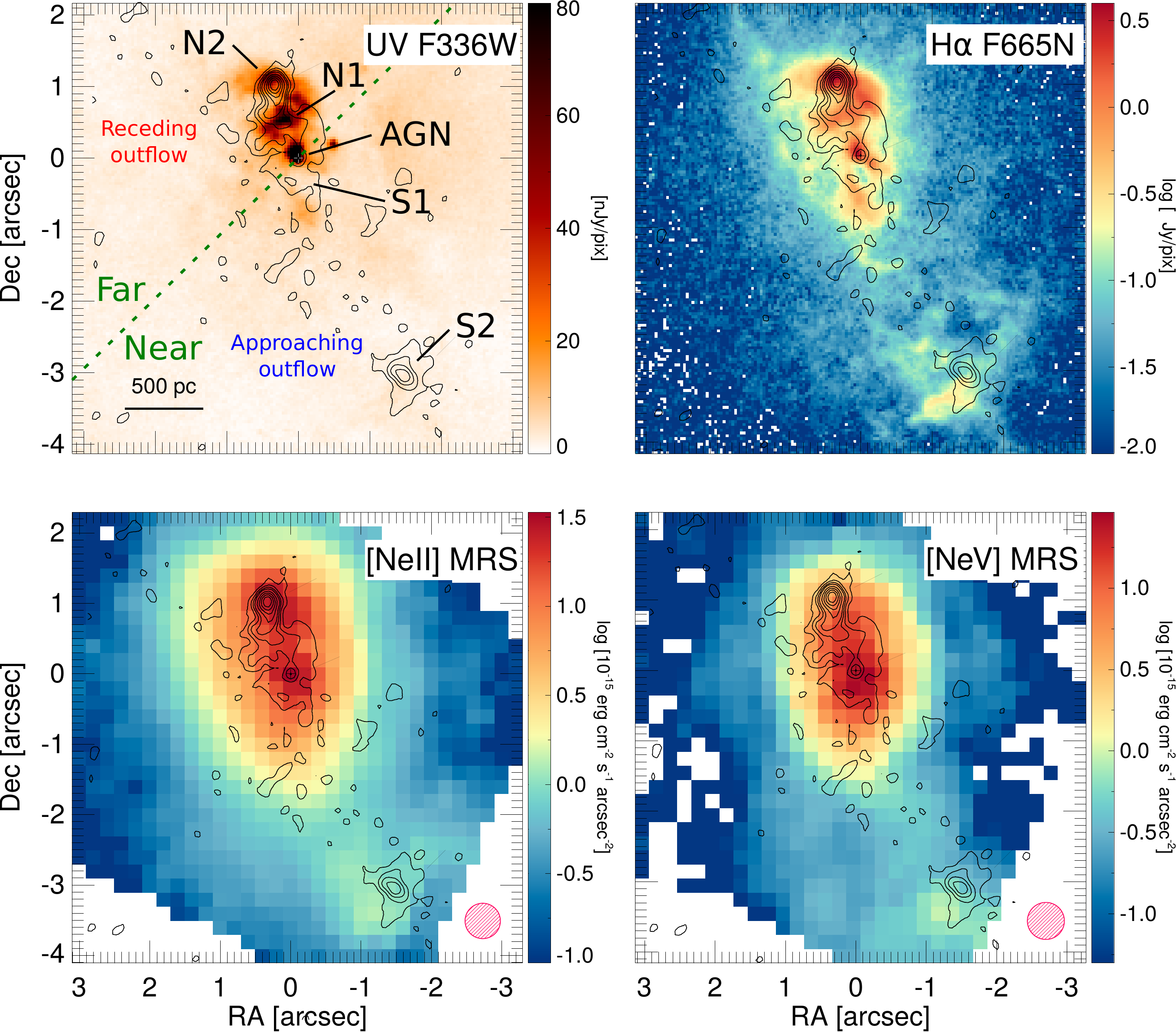}
\caption{NGC~7319 near-UV/blue image obtained with the F336W ($\lambda_{\rm p}$=0.335\micron) filter of {HST}\slash WFC3 (top left); H$\alpha$ emission from the continuum-subtracted {HST}\slash WFC3 F665N image (top right; Appendix~\ref{ss_ancillary}). The bottom panels show the {JWST}\slash MIRI MRS \Neii\ (left) and \Neva\ (right) emission line maps (zeroth-moment map). The contours are the MERLIN 1.4\,GHz (20\,cm) radio emission map from Fig.~7 of \citealt{Xanthopoulos2004}. The location of the radio hotspots, N2 and S2, and the nuclear diffuse radio lobes, N1 and S1, is indicated in the top-left panel. The dashed green line marks the stellar kinematic major axis \citep{Yttergren2021}. The far and near sides of the stellar disk are indicated (see Sect.~\ref{s_jet_ism}). The location of the receding (redshifted) and approaching (blueshifted) sides of the ionized outflow are indicated \citep{RodriguezBaras14, Yttergren2021}.
The red hatched circles represent the MIRI MRS PSF FWHM. \label{fig_mrs_system}}
\end{figure*}

\section{Introduction}\label{sec:intro}

\textit{James Webb} Space Telescope ({JWST}) observations of NGC~7319 were obtained as part of the Early Release Observations (ERO; Program ID \#2732; {\citealt{Pontoppidan2022ERO}}).
In this Letter we focus on the analysis of the Mid-Infrared Instrument (MIRI) medium resolution spectrograph (MRS) data \citep{Rieke2015, Wells2015, Wright2015}.

NGC~7319 is the largest galaxy in the Stephan's Quintet interacting group (Arp~319, HCG~092; $d=98$\,Mpc). NGC~7319 lost most of its atomic neutral gas during the interactions \citep{Williams2002}, but some atomic ionized and cold molecular gas is still present in its nuclear region \citep{RodriguezBaras14, Gao2000}.
This galaxy hosts a type 2 active galactic nucleus (AGN), which dominates the energy output of this object since no strong nuclear starburst signatures are found \citep{Sulentic2001}. 
From X-ray spectral fitting in \citet{Ricci2017_BAT}, the intrinsic hard X-ray (14--195\,keV) luminosity of 10$^{43.8}$\,erg\,s$^{-1}$ places the AGN in the range of local Seyferts, while the high column density ($N_{\rm H}$=10$^{23.8}$\,cm$^{-2}$) is consistent with its obscured nature.
Two asymmetric radio lobes and a compact core are detected, suggesting the presence of a low-power radio jet with $ L_{\rm 1.4\,GHz}$=10$^{22.5}$ W\,Hz$^{-1}$ \citep{Aoki1999, Xanthopoulos2004}.

Low-power radio jets produced by AGN seem to have an important role in galaxy evolution, providing kinetic feedback to the host interstellar medium (ISM) and regulating the formation of stars (e.g., \citealt{Weinberger2017, Dave2019}). 
The jet interaction with the cold molecular and ionized ISM phases has been studied in a few local galaxies and AGN with low $L_{\rm AGN}$ and $L_{\rm 1.4\,GHz}$, similar to NGC~7319, using submillimeter and optical data (e.g., \citealt{Alatalo2011, GarciaBurillo2014, GarciaBurillo2019, AlonsoHerrero2018, FernandezOntiveros2020, GarciaBernete2021, Venturi2021}). 
Previous mid-IR spectroscopic studies, however, focused on more luminous radio galaxies (e.g., \citealt{Ogle2010, Guillard2012, Dasyra2014, Zakamska2016}) and were unable to provide spatially resolved information.
In kinetic feedback studies, mid-IR emission lines are particularly valuable because {they can trace warm molecular} gas ($\sim$100--1000\,K) to highly ionized gas (coronal lines) and are less subject to extinction effects than optical/UV lines.
In this Letter we analyze the low-power jet--ISM interaction in NGC~7319 using MIRI\slash MRS data, which provides, for the first time, spatially resolved morphologies and kinematics.

\section{Data reduction and analysis}\label{s:data}

We processed the MIRI\slash MRS observations using the JWST calibration pipeline to produce 12 spectral cubes with the default spatial and spectral sampling. Technical details on the data reduction are described in Appendix~\ref{apx_reduction}.

\begin{figure*}
\centering
\includegraphics[width=0.98\textwidth]{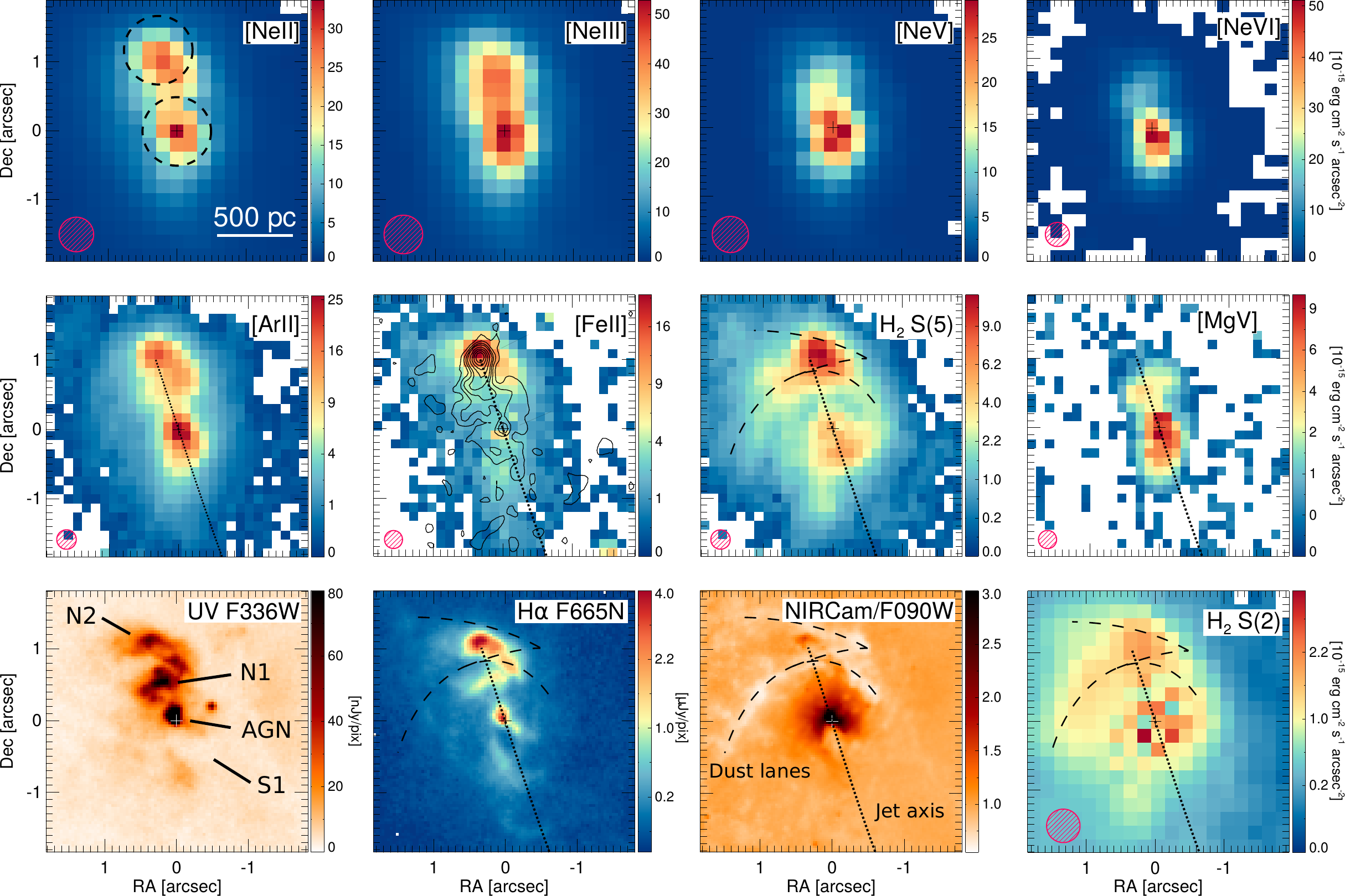}
\caption{Central 3\farcs6$\times$3\farcs6 region of NGC~7319, which includes the AGN and the northern radio lobe and hotspot N2. First row, from left to right: \Neii, \Neiii, \Neva, and \Nevi\ MIRI MRS line maps. The dashed circles on the [\ion{Ne}{ii}] map are the AGN and N2 extraction apertures.
Second row, from left to right: \Arii, \Feii, H$_2$ S(5) 6.91\micron, and \Mgv\ MIRI MRS line maps. The contours on the second panel represent the 1.4\,GHz  emission, as in Fig.~\ref{fig_mrs_system}. Third row: The first two panels are the same as the top-row panels of Fig~\ref{fig_mrs_system}. The third panel is the JWST/NIRCam F090W ($\lambda_{\rm p}$=0.90\micron) image after dividing by a Gaussian (FWHM$=$1\farcs2) smoothed version of itself to highlight finer details. The fourth panel is the H$_2$ S(2) 12.28\micron\ emission line. The dotted and dashed black lines trace the jet axis (N2 hotspot--AGN axis) and disk dust lanes, respectively. The red hatched circles represent the PSF FWHM, $\sim$0\farcs26--0\farcs60 depending on the wavelength, estimated from the unresolved AGN continuum. \label{fig_mrs_N2}}
\end{figure*}

We obtained maps of the brightest emission lines in this spectral range (Table~\ref{tbl_miri_lines}) by subtracting the local continuum (estimated with a linear fit) and integrating the line flux spaxel by spaxel between $\pm$1100\,km\,s$^{-1}$, which is the velocity range where emission is detected for the broader ionized gas lines. We also measured the first moment (velocity field) and second moment (velocity dispersion, $\sigma$) of the line emission profile. Line maps are presented in Figs.~\ref{fig_mrs_system}-\ref{fig_mrs_S2}.

We extracted the spectra of three regions (AGN, N2, and S2; Figs.~\ref{fig_mrs_system}, \ref{fig_mrs_N2}, and \ref{fig_mrs_S2}) using a 1\arcsec\ ($\sim$450\,pc) diameter aperture, which is $\sim$2 times the channel 3 point spread function (PSF) full-width half-maximum (FWHM). We applied a point-source correction to the AGN spectrum (see Appendix~\ref{apx_reduction}).
For N2 and S2, we did not apply any aperture correction since the emission appears more extended. The spectra are presented in Fig.~\ref{fig_mrs_spec}.

In addition, we used JWST and \textit{Hubble} Space Telescope (HST) imaging data to trace the UV, H$\alpha$, and dust lane morphologies (see Appendices~\ref{ss_ancillary} and \ref{s_astrometry}).

\section{Jet--ISM interaction in NGC~7319}\label{s_jet_ism}

Interferometric radio observations at 1.4, 5, and 8\,GHz revealed an unresolved core, with a flat radio spectrum, and two synchrotron hotspots at the ends of the radio lobes. The hotspots are asymmetrically located at 430\,pc and 1.5\,kpc projected distances, respectively, from the core, which implies a high arm-length ratio of $\sim$3.5 (Fig.~\ref{fig_mrs_system} and \citealt{Aoki1999, Xanthopoulos2004}).
Relativistic beaming can produce asymmetric arm-length ratios.
Assuming intrinsically symmetric jet propagation, the approaching side is brighter and its apparent distance higher \citep{Boettcher2012}. On the contrary, in NGC~7319 the brighter side is closer to the core and coincident {with the receding} side of the ionized outflow (see below and Fig.~\ref{fig_mrs_system}).
Alternatively, the length asymmetry can be produced by the interaction of the northern radio jet with the ISM (e.g., \citealt{Xanthopoulos2004}).

This galaxy contains a $\sim$400\,km\,s$^{-1}$ ionized gas outflow, detected in [\ion{O}{iii}]5007\AA\ and H$\alpha$. This outflow is co-spatial with the radio lobes, and its kinematic major axis is aligned with the core-hotspots axis (see \citealt{Yttergren2021}), which suggests that the outflow is driven by the radio jet \citep{Aoki1996}.

Assuming that the large-scale spiral pattern and tidal features of NGC~7319 resulting from prograde galaxy interactions are trailing \citep{Renaud2010}, the stellar disk is rotating counterclockwise. This assumption, together with the stellar kinematics \citep{Yttergren2021}, determines the far and near sides of the disk (see the top-left panel of Fig.~\ref{fig_mrs_system}). Therefore, the northern side of the jet (receding outflow) is seen in projection on the far side of the disk, while the southern side (approaching outflow) is over the near side of the disk. This geometry and outflow kinematics is at odds with a perpendicular-to-the-disk jet orientation. Instead, it is consistent with the jet axis being close to the plane of the disk, and, therefore, it favors jet--ISM interactions.

The top panels of Fig.~\ref{fig_mrs_system} show the near-UV/blue continuum and H$\alpha$ emissions, which are spatially coincident with the radio lobes and peak close to the northern hotspot, N2. Bow shocks are also present at both the northern and southern lobes, especially in the H$\alpha$ image. The bottom panels show the MRS maps of the \Neii\ (ionization potential, IP, = 22\,eV) and the high-ionization \Neva\ (IP = 97\,eV) emission lines. 
Both mid-IR lines have morphologies that follow the radio jet axis and resemble that of the higher angular resolution UV/optical imaging data. So they are likely connected to the jet as well. 
Nevertheless, a contribution to these lines from the AGN extended narrow line region is also possible.

\setcounter{figure}{3}
\begin{figure*}
\centering
\includegraphics[width=0.98\textwidth]{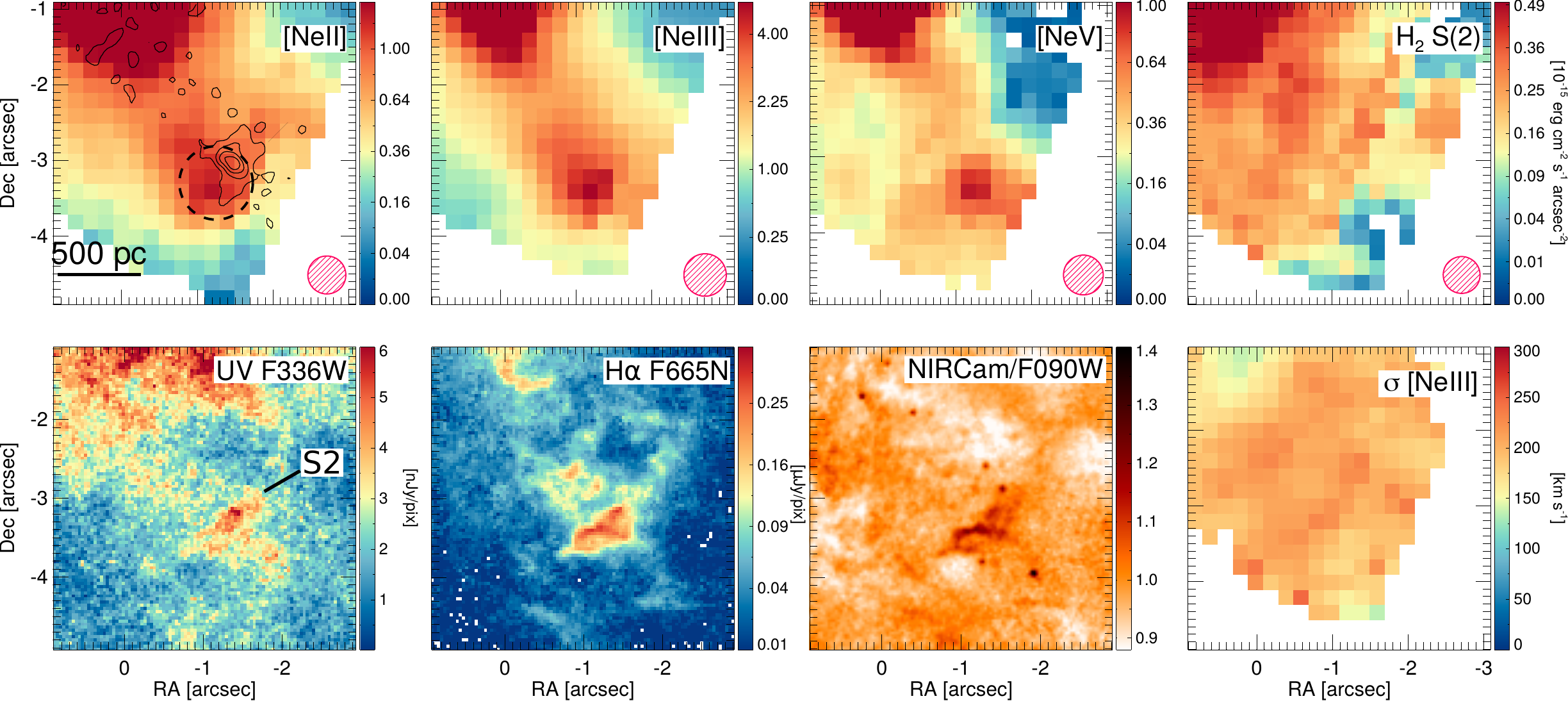}
\caption{Southern radio lobe and hotspot S2 of NGC~7319. First row, from left to right: \Neii, \Neiii, \Neva, and H$_2$ S(2) 12.28\micron\ MIRI MRS line maps. The contours on the first panel represent the 1.4\,GHz  emission for reference, as in Fig.~\ref{fig_mrs_system}. The dashed circle on the [\ion{Ne}{ii}] map is the S2 extraction aperture.
The red hatched circles represent the PSF FWHM, $\sim$0\farcs49--0\farcs56.
Second row: The first three panels are the same as the last-row panels of Fig~\ref{fig_mrs_N2}. The fourth panel is the velocity dispersion of the \Neiii\ emission line. \label{fig_mrs_S2}}
\end{figure*}

\setcounter{figure}{2}
\begin{figure}
\centering
\includegraphics[width=0.48\textwidth]{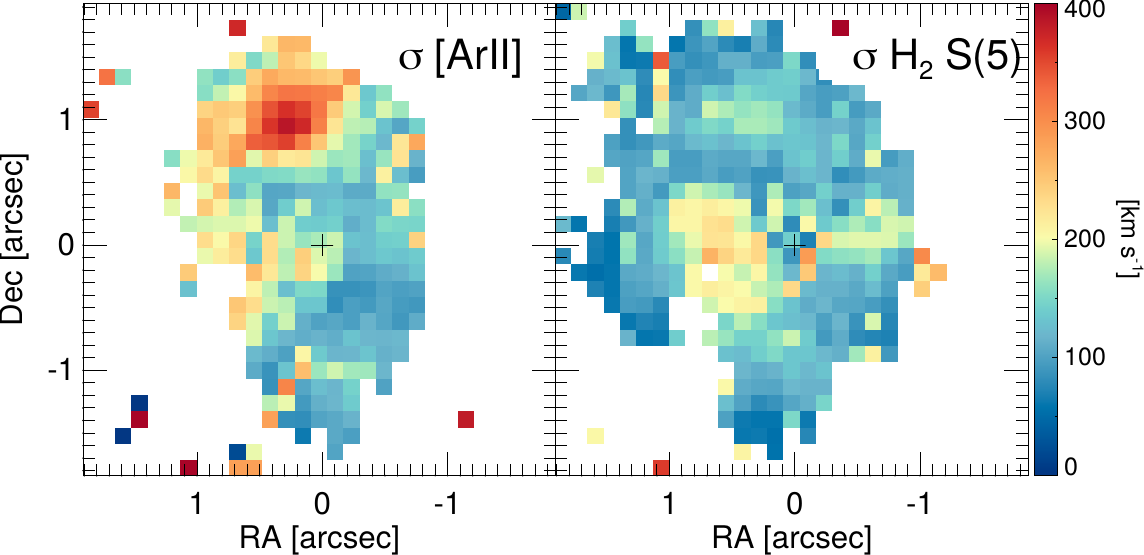}
\caption{Velocity dispersion of the \Arii\ (left) and H$_2$ S(5) 6.91\micron\ (right) emission lines. The FoV is the same as in Fig.~\ref{fig_mrs_N2}. The velocity field is shown in Fig.~\ref{fig_mrs_N2_v}. \label{fig_mrs_N2_sigma}}
\end{figure}

\setcounter{figure}{4}

\subsection{Jet--molecular gas interaction: Northern radio hotspot}

Figure~\ref{fig_mrs_N2} shows the line maps centered on the more luminous northern radio lobe. The first row compares the Ne emission from Ne$^+$ to Ne$^{5+}$ (IP=126\,eV). These maps trace the ionization state of the gas, avoiding uncertainties due to relative atomic abundances.
We find that the extent of the Ne emission decreases with increasing ionization stage.
The lower-ionization [\ion{Ne}{ii}] emission peaks close to N2, with reduced emission at N1 closer to the nucleus. Intermediate [\ion{Ne}{iii}] emission seems to peak between N1 and N2, while the highest-ionization [\ion{Ne}{v}] and [\ion{Ne}{vi}] lines, instead, are brighter at N1. The latter indicates that the intense UV radiation from N1 is unlikely produced by star formation.
To examine the excitation at higher angular resolution, we used the short wavelength \hbox{channel 1} \Arii\ (IP=15.8\,eV) and \Mgv\ (IP=109\,eV) emission lines as proxies of the low- and high-ionization gas 
(second row of Fig.~\ref{fig_mrs_N2}). The better angular resolution allows us to establish that the high-ionization emission is mostly concentrated at N1 to the north and at S1 to the south of the nucleus, while lower-ionization gas dominates the emission at the N2 radio hotspot. The [\ion{Ar}{ii}] morphology is also similar to that of the H$\alpha$ map (bottom row).

Radio jets interact with the ISM, transferring energy and momentum to the ISM clouds through shock waves \citep{Sutherland2007}. These shock waves enhance the mid-IR rotational H$_2$ emission in radio galaxies \citep{Guillard2012, Ogle2010}. In NGC~7319, the H$_2$ S(5) 6.91\micron\ emission peaks close to the N2 hotspot. 
Both H$_2$ S(2) 12.28\micron\ and H$_2$ S(5) also trace the dust lanes visible in the F090W image (see Fig.~\ref{fig_mrs_N2}). Therefore, the radio hotspot seems to be located $<$100\,pc away from the intersection between the dust lanes and the jet axis.
We note, however, that the contrast between the dust lanes and hotspot emission is smaller for the lower excitation H$_2$ S(2) transition.
The jet--ISM interaction at N2 is also supported by the bright \Feii\ emission, which is enhanced in shocks {\citep{Allen2008, Koo2016}}.

The polycyclic aromatic hydrocarbon (PAH) emission at N2 is weak relative to the H$_2$ lines (Fig.~\ref{fig_mrs_spec}), as seen in shock-excited regions \citep{Guillard2012, Beirao2015}. {The ratio between the 11.3\micron\ PAH flux, estimated by subtracting a local continuum and integrating the spectrum between 11.0 and 11.8\micron, and the H$_2$ S(1) line flux is $\sim$2.5}, which is $\sim$10 times lower than in local starburst galaxies (see \citealt{Pereira2010}). Therefore, photodissociation region excitation at N2 is unlikely.
In addition, the 7.7\micron\ PAH feature is extremely weak at N2 (Fig.~\ref{fig_mrs_spec}), and only the 11.3\micron\ PAH, which is more resilient in hard environments \citep{GarciaBernete2022}, is clearly detected. Similar spectra are seen in radio galaxies with strong jets, which differ from star-forming galaxies where the 7.7\micron\ PAH feature is dominant \citep{Ogle2010, Zakamska2016, Smith07}.

The ionized gas velocity dispersion (turbulence) is increased at N2 ($\sigma$\,$\sim$300\,km\,s$^{-1}$). The $\sigma$ of the rotational H$_2$ transitions ($\sim$150\,km\,s$^{-1}$), however, is not enhanced  (Fig.~\ref{fig_mrs_N2_sigma} and Table~\ref{tbl_miri_lines}), which is consistent with observations of radio galaxies \citep{Guillard2012}.

Low-power jets ($P_{\rm jet}<$10$^{43}$\,erg\,s$^{-1}$) cannot easily pierce dense gas clouds and might remain trapped in the ISM \citep{Mukherjee2016}. We estimate $P_{\rm jet}\sim$2$\times$10$^{43}$\,erg\,s$^{-1}$ (Sect.~\ref{s:mass}), so this could be the case for the northern {lobe if decelerated by molecular clouds} in the disk.

\subsection{Jet--atomic gas interaction: Southern radio hotspot}

The more distant southern lobe was only observed with the longer wavelength MRS channels 3 and 4 due to the larger field of view (FoV) at those wavelengths. For this reason, a reduced set of mid-IR lines is available (Fig.~\ref{fig_mrs_S2}). 
The three Ne lines peak at the same location close to the S2 radio hotspot, although, we note that the offset between the Ne and radio peaks is larger at S2 than at N2 (0\farcs2 vs. 0\farcs5). In the higher spatial resolution {HST} and JWST\slash Near Infrared Camera (NIRCam) continuum and H$\alpha$ images, this region appears as an arc or bow-shock. This structure was already identified by \citet{Aoki1999}. The velocity dispersion of the ionized gas is high, $\sim$220--300\,km\,s$^{-1}$ (Table~\ref{tbl_miri_lines}), and similar to that of N2, although the S2 hotspot does not stand out in the velocity dispersion map (last panel of Fig.~\ref{fig_mrs_S2}).
Contrary to the N2 hotspot, at S2 the H$_2$ emission is not enhanced, although in this case the S(5) line, more sensitive to the jet-excited molecular gas, is not available. The higher H$_2$ S(1) to S(2) ratio in this region also supports a lower excitation of the molecular gas in S2 (Table~\ref{tbl_miri_ratios}). Conversely, the higher [\ion{Ne}{iii}] and [\ion{Ne}{v}] to [\ion{Ne}{ii}] ratios indicate that the ionized gas phase is more highly ionized in S2 (Table~\ref{tbl_miri_ratios}).
These results suggest that the jet is interacting more strongly with the atomic gas phase, relative to the molecular, at S2 than at N2.

\begin{table}
\caption{MIRI\slash MRS line ratios.}
\label{tbl_miri_ratios}
\centering
\begin{small}
\begin{tabular}{lcccccccccc}
\hline \hline
\\
Ratio & AGN & N2 & S2 \\
\hline
{[\ion{Ne}{iii}]}\slash{[\ion{Ne}{ii}]} & 2.08 & 1.16 & 3.31 \\
{[\ion{Ne}{v}]}\slash{[\ion{Ne}{ii}]} & 0.96 & 0.22 & 0.60 \\
{[\ion{Ne}{v}]}\slash{[\ion{Ne}{iii}]} & 0.46 & 0.19 & 0.18 \\
{[\ion{Ne}{v}]}\slash{H$_2$ S(1)} & 10.6 & 1.85 & 1.43 \\
{H$_2$ S(1)}\slash{H$_2$ S(2)} & 0.92 & 1.27 & 2.70 \\

\hline
\end{tabular}
\end{small}
\tablefoot{Line ratios between the \Neii, \Neiii, \Neva, H$_2$ S(1), and H$_2$ S(2) emission lines. All the lines are observed with {channel 3 of the MRS}. The estimated ratio uncertainty, based on the flux uncertainties, is $\sim$10\% (see Table~\ref{tbl_miri_lines}).
AGN Ne line ratios are similar to the median ratios measured in local Seyfert galaxies \citep{Pereira2010c}. 
}
\end{table}

\section{Ionized and warm\slash hot molecular gas mass affected by the jet}\label{s:mass}

We derived the ionized gas mass using Eq.~1 of \citet{Venturi2021}. The ratio between H$\alpha$ and \ion{H}{i} 6--5 7.46\micron\ (Pf$\alpha$) is 112 at 10\,000\,K \citep{Storey1995}. 
We assumed that the electron density, $n_{\rm e}$, in the ionized gas in the shock is between 100\,cm$^{-3}$ and the upper limit, $n_{\rm e}$<500\,cm$^{-3}$, obtained from the \Feii\slash [\ion{Fe}{ii}]4.89\micron\ $>$60 ratio (Fig.~\ref{fig_fe2_density}). The resulting mass range is $M_{\rm ion} = $ (2.4--12)$\times$10$^5$\,\Msun.

\begin{figure}
\centering
\vspace{5mm}
\includegraphics[width=0.35\textwidth]{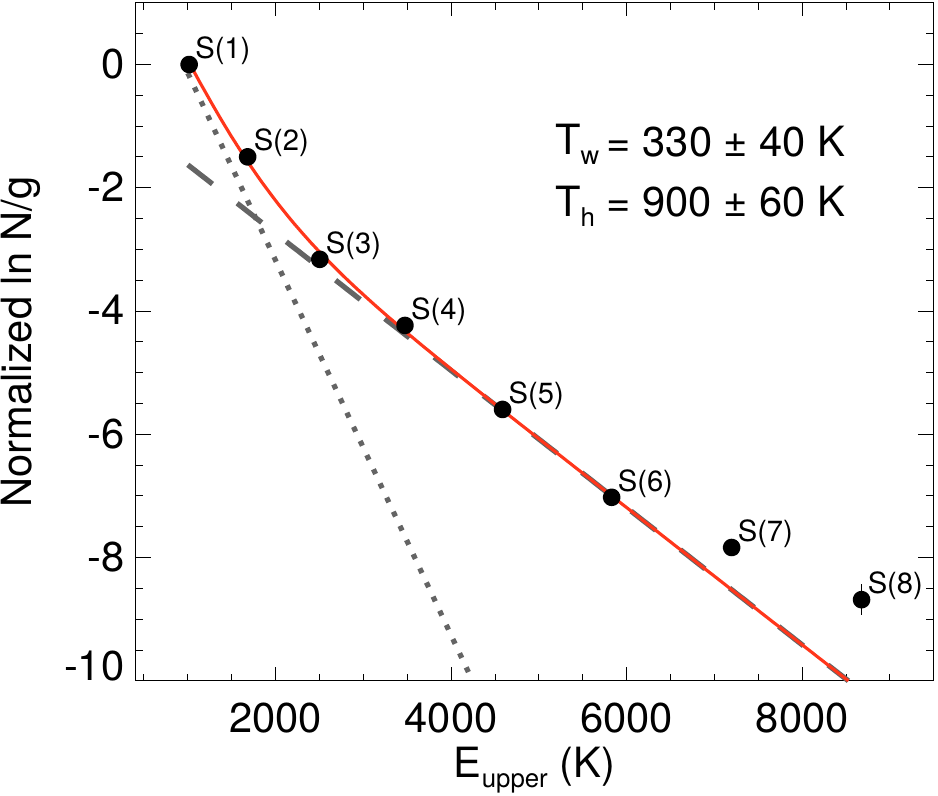}
\caption{Rotational diagram showing the H$_2$ 0--0 S(1) to S(8) transitions measured at the northern radio hotspot, N2, assuming the ortho-to-para ratio of 3 expected for $T_{\rm rot}$>200\,K \citep{Burton1992}. 
The red line is the best LTE fit with two temperatures to the S(1)-S(6) lines. The dotted (dashed) line represents the warm (hot) component. The solid red line is the sum of the two components.
\label{fig_h2_rotd}}
\end{figure}

The warm and hot molecular gas mass at N2 is derived using the H$_2$ S(1) to S(8) transitions. 
The critical densities of these transitions are relatively low, $n_{\rm H_2}$ = 10$^2$ to 10$^5$\,cm$^{-3}$ at 500\,K \citep{LeBourlot1999}, so we can assume local thermodynamic equilibrium (LTE) conditions. 
Under this assumption, the S(1) to S(6) transitions can be fit with a two-temperature model: a warm component with $T_{\rm w}$=330$\pm$40\,K and $M_{\rm H_2,w}$=(6.0$\pm$1.4)$\times$10$^5$\,\Msun, which dominates the S(1) and S(2) emission; and a hotter component with $T_{\rm h}$=900$\pm$60\,K and $M_{\rm H_2,h}$=(0.44$\pm$0.12)$\times$10$^5$\,\Msun, which dominates the S(3)--S(6) emission (Fig.~\ref{fig_h2_rotd}).
The higher $J$ transitions S(7) and S(8) deviate from this fit, suggesting the existence of a hotter component, whose temperature is $>$1000\,K but not well constrained by these transitions alone. 
Molecular gas is expected to reform in post-shock regions \citep{Hollenbach1989}. However, the lower H$_2$ velocity dispersion compared to that of the ionized gas (Fig.~\ref{fig_mrs_N2_sigma}) suggests that the origin of the H$_2$ emission is not reformed H$_2$ molecules after the cooling of that ionized gas. Instead, it is plausible that this H$_2$ was already present and is being excited by the shock waves produced by the jet.

The ionized gas velocity dispersion is $\sim$2 times higher than that of the warm H$_2$, and the $M_{\rm ion}$\slash $M_{\rm H_2,w}$ ratio is $\sim$0.3--2.5. Therefore, the mechanical energy of the ionized phase, $E_{\rm ion}$={3\slash 2} $M_{\rm ion}\sigma^2$=(0.6--3.2)$\times$10$^{54}$\,erg, is $\sim$1.3--10 times higher than that of the warm H$_2$, $E_{\rm H_2,w}$=(0.4$\pm$0.1)$\times$10$^{54}$\,erg.

We estimated a jet cavity power of $P_{\rm jet}$ $\sim$2$\times$10$^{43}$\,erg\,s$^{-1}$ using Eq.~16 of \citet{Birzan2008} and $L_{\rm 1.4\,GHz}$=3.3$\times$10$^{22}$\,W\,Hz$^{-1}$ \citep{Aoki1999}, which is similar to NGC~1068 \hbox{\citep{GarciaBurillo2014}}, {and we assume that half of it reaches the northern radio hostspot N2.}
The ratio ($E_{\rm H_2,w}$+$E_{\rm ion})$\slash ($t_{\rm jet} P_{\rm jet}\slash 2$) is $\sim$(2.4--9.7)$\times$10$^{-3}$, where the jet life, $t_{\rm jet}$\,$\sim$1.1$\times$10$^{6}$\,yr, is estimated from the jet travel time to S2 following \citet{Venturi2021}.
This indicates that only a small fraction of the jet energy {($<$1\%)} remains as mechanical energy in the ionized and warm molecular ISM phases. 

\section{Summary and conclusions}

We have analyzed the low-power ($L_{\rm 1.4\,GHz}$=3.3$\times$10$^{22}$\,W\,Hz$^{-1}$) jet--ISM interaction in the Seyfert 2 galaxy NGC~7319 using \JWST\slash MIRI MRS data. We find evidence suggesting that molecular gas in dust lanes decelerates the jet at the northern hotspot, N2, which is three times closer to the nucleus than the southern hotspot, S2.
Enhanced warm and hot H$_2$ emission ($T_{\rm H_2}$\,$\sim$330\,K and 900\,K) and ionized gas tracers, as well as \Feii\ emission, which is a shock tracer, are detected at N2 at the intersection between the jet axis and dust lanes.
On the contrary, at the more distant S2 hotspot, the H$_2$ excitation is lower and the atomic gas is more highly ionized. This suggests that the jet interacts more strongly with the atomic gas there.
Therefore, the reduced molecular gas--jet interaction at the southern radio lobe could make it easier for the jet to reach greater distances.

Extended (distances of up to 1.5\,kpc from the nucleus) high-ionization (\Mgv, \Nevi, \Neva, and \Nevb) emission is detected close to the radio jet hotspots. These lines can be produced in the photoionized precursor of the shock waves produced by the jet or {can be gas photoionized} by the AGN radiation in an extended narrow line region.

At N2, $M_{\rm ion}$=(2.4--12)$\times$10$^5$\,\Msun\ is comparable to the warm ($\sim$330\,K) H$_2$ gas mass and $>$6 times higher than the hot ($\sim$900\,K) H$_2$ mass. The ionized gas is also more turbulent ($\sigma_{\rm ion}\sim300$ vs. $\sigma_{\rm H_2}\sim150$\,km\,s$^{-1}$). Therefore, the mechanical energy of the ionized gas is $\sim$1.3--10 times higher than that of the warm$+$hot molecular gas. From these estimates, we find that {$<$1\%} of the jet energy remains as mechanical energy in these two ISM phases, {which is much lower than the 25--30\%\ of the jet energy that is injected into the ISM according to simulations \citep{Mukherjee2016}.}
{We note that the jet can also transfer mechanical energy to the cold molecular gas traced by CO and launch molecular outflows (e.g., \citealt{Morganti2015, RamosAlmeida2022}).}

These initial results show the \JWST\slash MIRI MRS capabilities to constrain the AGN--jet feedback on the multiphase ISM at unprecedented sensitivity and unprecedented angular and spectral resolutions in the mid-IR.

\begin{acknowledgements}
{We thank the referee for their useful comments and suggestions.}
We thank P. Gandhi and E. Hicks for the careful reading of the manuscript and useful discussion.
The authors acknowledge the ERO team for developing their observing program with a zero-exclusive-access period.

MPS acknowledges support from the Comunidad de Madrid through the Atracci\'on de Talento Investigador Grant 2018-T1/TIC-11035 and PID2019-105423GA-I00 (MCIU/AEI/FEDER,UE).
JAM, AL, and LC acknowledge support by grant PIB2021-127718NB-100 by the Spanish Ministry of Science and Innovation/State Agency of Research (MCIN/AEI).
IGB acknowledges support from STFC through grant ST/S000488/1.
AAH acknowledges support from grant PGC2018-094671-B-I00 funded by MCIN/AEI/ 10.13039/501100011033 and by ERDF A way of making Europe.
SGB  acknowledges support from the research project PID2019-106027GA-C44 of the Spanish Ministerio de Ciencia e Innovaci\'on.
CRA acknowledges support from grant PID2019-106027GB-C42, funded by MICINN-AEI/10.13039/501100011033, from EUR2020-112266, funded by  MICINN-AEI/10.13039/501100011033 and the European Union NextGenerationEU/PRTR, and from the Consejer\' ia de Econom\' ia, Conocimiento y Empleo del Gobierno de Canarias and the European Regional Development Fund (ERDF) under grant ProID2020010105, ACCISI/FEDER, UE.

This work is based on observations made with the NASA/ESA/CSA James Webb Space Telescope. The data were obtained from the Mikulski Archive for Space Telescopes at the Space Telescope Science Institute, which is operated by the Association of Universities for Research in Astronomy, Inc., under NASA contract NAS 5-03127 for JWST; and from the European JWST archive (eJWST) operated by the ESAC Science Data Centre (ESDC) of the European Space Agency. These observations are associated with program \#2732, \#1049, \#1050.

This research is based on observations made with the NASA/ESA Hubble Space Telescope obtained from the Space Telescope Science Institute, which is operated by the Association of Universities for Research in Astronomy, Inc., under NASA contract NAS 5–26555. These observations are associated with programs 11502 and 12301.

\end{acknowledgements}

\appendix

\section{Data reduction}

\subsection{MIRI\slash MRS data reduction}\label{apx_reduction}

MIRI\slash MRS covers the mid-IR spectral range between 4.9 and 28.1\micron. This spectral range is split into four channels (channel 1 from 4.9 to 7.65\micron; channel 2 from 7.51 to 11.71\micron; channel 3 from 11.55 to 18.02\micron; and channel 4 from 17.71 to 28.1\micron). Each channel is in turn divided into three sub-bands (short, medium, and long), which each cover a third of the channel spectral range. A single exposure simultaneously observes a single sub-band for the four channels. The FoV and pixel size are smaller for the shorter wavelength channels (from a 3\farcs2$\times$3\farcs7 FoV and $\sim$0\farcs2 pixel in channel 1 to a 6\farcs6$\times$7\farcs6 FoV and $\sim$0\farcs3 pixel in channel 4) to better sample the diffraction-limited JWST PSF (see \citealt{Rieke2015, Wells2015, Wright2015}).

We downloaded the uncalibrated NGC~7319 data from the JWST archive. The MIRI\slash MRS observations were processed using the JWST calibration pipeline (release 1.6.3) with context 0939 of the Calibration References Data System (CRDS). We followed the standard pipeline procedure to generate the fully calibrated detector products (level 2b) for the on-source and background observations (see \citealt{AlvarezMarquez2022} for a more comprehensive explanation of the MRS calibration data process). Before generating the three-dimensional MRS spectral cubes, we applied a residual fringe correction in the detector plane\footnote{\href{https://jwst-pipeline.readthedocs.io/en/latest/jwst/residual_fringe/index.html}{https://jwst-pipeline.readthedocs.io/en/latest/jwst/residual\_fringe/\\index.html.}}. This step corrects the low frequency fringe residuals remaining after the standard pipeline fringe flat correction (\citealt{Argyriou2020}, Gasman et al. in prep., Kavanagh et al. in prep.).

Then, we generated 12 three-dimensional spectral cubes (the three sub-bands -- short, medium, and long bands -- for the four MRS channels) with the default spatial and spectral samplings.
We estimated the background emission by calculating the median value for each wavelength channel in the background data cubes. This median value for each spectral channel {was subtracted from} the on-source data cubes.

For the nuclear spectra of the AGN, an additional residual fringe correction was implemented to correct for the high frequency fringes generated in the dichroics, which are noticeable in the spectra of bright point sources in channels 3 and 4. With these final corrections, the final fringe residuals are reduced to levels lower than 6\%, with a median level of 2-4\%\ (Kavanagh et al. in prep.).

From the emission line fits, we estimate that the current MRS wavelength calibration has, in general, offsets smaller than $\pm$100\,km\,s$^{-1}$, consistent with the JWST commissioning report \citep{Rigby2022}.\ The exceptions are the medium sub-band of channel 3 (13.29--15.52\micron), where a constant shift of $\sim$0.04\micron\ ($\sim$900\,km\,s$^{-1}$) is present, and the medium sub-band of channel 2 (8.67--10.15\micron), where the wavelength shift is variable. In addition, the wavelength solution of the long sub-band of channel 3 around $\sim$17.3\micron\ (observed wavelength of the H$_2$ S(1) transition in NGC~7319) is affected by artifacts that alter the line profile by creating spurious double-peaked profiles at some positions of the data cube but they do not alter the integrated line flux.
These issues have been identified and do not affect the analysis presented in this work.

The unresolved line FWHM ($c$\slash resolving power) increases from $\pm$80\,km\,s$^{-1}$ to 130\,km\,s$^{-1}$ with increasing wavelength for the channels analyzed here (1, 2, and 3; \citealt{Labiano2021}, Jones et al. in prep.). Therefore, all the emission lines in NGC~7319, which have FWHM\slash 2.35=$\sigma$=130--370\,km\,s$^{-1}$ (Table~\ref{tbl_miri_lines}), are spectrally resolved.

We estimated the point-source correction for a 1\arcsec\ diameter aperture using the MRS observations of HD~163466 (Program ID \#1050) for channels 1, 2, and 3, and SMP-LMC-058 (Program ID \#1049) for channel 4.

\subsection{Ancillary JWST and \HST\ imaging }\label{ss_ancillary}

JWST/NIRCam and MIRI wide-band filter images are also available as part of the ERO data release. 
We obtained the calibrated level 3 images (pipeline release 1.5.3 and context 0919 of the CRDS) from the JWST archive.
We used the NIRCam F090W ($\lambda_{\rm p}$=0.90\micron; $\Delta \lambda$=0.19\micron)\footnote{\href{https://jwst-docs.stsci.edu/jwst-near-infrared-camera/nircam-instrumentation/nircam-filters}{https://jwst-docs.stsci.edu/jwst-near-infrared-camera/nircam-instrumentation/nircam-filters}}, and the MIRI F1000W ($\lambda_{\rm p}$=10.0\micron; $\Delta \lambda$=2.0\micron)\footnote{\href{https://jwst-docs.stsci.edu/jwst-mid-infrared-instrument/miri-instrumentation/miri-filters-and-dispersers}{https://jwst-docs.stsci.edu/jwst-mid-infrared-instrument/miri-instrumentation/miri-filters-and-dispersers}} images, which we used to determine the coordinates of the AGN mid-IR emission (Appendix~\ref{s_astrometry}).

\textit{Hubble} Space Telescope Wide Field Camera 3 (WFC3) images were also retrieved from the Hubble Legacy Archive: the F336W ($\lambda_{\rm p}$=0.335\micron; $\Delta \lambda$=0.051\micron)\footnote{\href{https://hst-docs.stsci.edu/wfc3ihb/chapter-6-uvis-imaging-with-wfc3/6-5-uvis-spectral-elements}{https://hst-docs.stsci.edu/wfc3ihb/chapter-6-uvis-imaging-with-wfc3/6-5-uvis-spectral-elements}}, which covers the near-UV/blue emission of the galaxy; and the F665N ($\lambda_{\rm p}$=0.666\micron; $\Delta \lambda$=0.013\micron), which includes the H$\alpha$ emission and the weakest line, 6548\AA, of the [\ion{N}{ii}] doublet at the redshift of NGC~7319. We subtracted the underlying continuum by linearly interpolating the F606W ($\lambda_{\rm p}$=0.59\micron; $\Delta \lambda$=0.22\micron) and F814W ($\lambda_{\rm p}$=0.80\micron; $\Delta \lambda$=0.16\micron) images at the mean wavelength of the F665N filter, taking into account that the F606W image includes the F665N emission.
The [\ion{N}{ii}]6548\AA\slash H$\alpha$ ratio is $<$0.3 in the central $\sim$10\arcsec, based on optical integral field spectroscopy \citep{RodriguezBaras14}, so the continuum-subtracted F665N morphology is likely dominated by the H$\alpha$ emission.

\subsection{Astrometric registration}\label{s_astrometry}
 
To allow for morphology comparisons between the MRS line maps and the JWST and HST ancillary images (Appendix~\ref{ss_ancillary}), we require relative astrometric differences smaller than 0.06\arcsec (channel 1 data cube half pixel).
To achieve this, we registered all the images and line maps to a common frame. We used as reference the NIRCam F090W image. 
The background quasar J223603.7$+$335824, which is located 8\arcsec south of the nucleus and is well detected in all the images, was used to align them. The mid-IR AGN coordinates were derived from the F1000W MIRI image. These coordinates were then assigned to the location of the mid-IR AGN continuum peak derived for each of the MRS line maps.

\onecolumn

\section{MIRI/MRS emission lines}

\begin{table*}[!h]
\caption{MIRI\slash MRS emission lines.}
\label{tbl_miri_lines}
\centering
\begin{tiny}
\begin{tabular}{lcccccccccccc}
\hline \hline
\\
Transition & $\lambda_{\rm rest}$ & IP\,\tablefootmark{a} & $\log n_{\rm crit}$\,\tablefootmark{b} & \multicolumn{2}{c}{AGN} & & \multicolumn{2}{c}{N2} & & \multicolumn{2}{c}{S2} \\
\cline{5-6} \cline{8-9} \cline{11-12}\\[-1.5ex]
&  & & & Flux & $\sigma$ & & Flux & $\sigma$ & & Flux & $\sigma$ \\
& \micron & eV & cm$^{-3}$ & 10$^{-16}$\,erg\,cm$^{-2}$\,s$^{-1}$ & km\,s$^{-1}$ & & 10$^{-16}$\,erg\,cm$^{-2}$\,s$^{-1}$ & km\,s$^{-1}$ & & 10$^{-16}$\,erg\,cm$^{-2}$\,s$^{-1}$ & km\,s$^{-1}$ \\

\hline
\multicolumn{13}{c}{MIRI\slash MRS Channel 1}  \\
\hline
{[\ion{Fe}{II}]} & 4.889 & 7.9 & 4.39 & $<$1.45 & \nodata  &  & $<$0.87 & \nodata  &  & \nodata  & \nodata  &  \\
H$_2$ 0--0 S(8) & 5.053 & \nodata & \nodata & 4.2 $\pm$ 1.1 & \nodata &  & 5.5 $\pm$ 1.2 & \nodata &  & \nodata  & \nodata  &  \\
{[\ion{Fe}{II}]} & 5.340 & 7.9 & 3.09 & 27.5 $\pm$ 1.4 & 253 $\pm$ 47 &  & 51.27 $\pm$ 0.44 & 336 $\pm$ 6 &  & \nodata  & \nodata  &  \\
{[\ion{Fe}{VIII}]} & 5.447 & {125} & 6.41 & 14.13 $\pm$ 0.63 & \nodata &  & $<$0.69 & \nodata  &  & \nodata  & \nodata  &  \\
{[\ion{Mg}{VII}]} & 5.503 & {187} & 6.53 & 19.5 $\pm$ 1.5 & \nodata &  & $<$0.70 & \nodata  &  & \nodata  & \nodata  &  \\
H$_2$ 0--0 S(7) & 5.511 & \nodata & \nodata & 19.1 $\pm$ 2.3 & \nodata &  & 19.02 $\pm$ 0.40 & 134 $\pm$ 16 &  & \nodata  & \nodata  &  \\
{[\ion{Mg}{V}]} & 5.610 & 109 & 6.60 & 34.5 $\pm$ 1.3 & 168 $\pm$ 49 &  & 3.43 $\pm$ 0.33 & \nodata &  & \nodata  & \nodata  &  \\
H$_2$ 0--0 S(6) & 6.109 & \nodata & \nodata & 6.11 $\pm$ 0.65 & \nodata &  & 6.58 $\pm$ 0.49 & \nodata &  & \nodata  & \nodata  &  \\
{[\ion{Ni}{II}]} & 6.636 & 7.6 & 5.92 & $<$0.68 & \nodata  &  & 2.74 $\pm$ 0.23 & 241 $\pm$ 69 &  & \nodata  & \nodata  &  \\
{[\ion{Fe}{II}]} & 6.721 & 7.9 & 3.09 & $<$1.70 & \nodata  &  & 1.79 $\pm$ 0.23 & 222 $\pm$ 67 &  & \nodata  & \nodata  &  \\
H$_2$ 0--0 S(5) & 6.909 & \nodata & \nodata & 34.8 $\pm$ 1.0 & 196 $\pm$ 37 &  & 33.05 $\pm$ 0.28 & 144 $\pm$ 17 &  & \nodata  & \nodata  &  \\
{[\ion{Ar}{II}]} & 6.985 & 15.8 & 5.62 & 87.1 $\pm$ 1.8 & 176 $\pm$ 24 &  & 48.97 $\pm$ 0.27 & 328 $\pm$ 3 &  & \nodata  & \nodata  &  \\
{[\ion{Na}{III}]} & 7.318 & 47.3 & 6.80 & 8.6 $\pm$ 1.1 & 321 $\pm$ 71 &  & 2.73 $\pm$ 0.18 & 367 $\pm$ 38 &  & \nodata  & \nodata  &  \\
\hline
\multicolumn{13}{c}{MIRI\slash MRS Channel 2}  \\
\hline
H\,I 6-5 & 7.460 & \nodata & \nodata & 7.2 $\pm$ 1.2 & \nodata &  & 2.89 $\pm$ 0.34 & 307 $\pm$ 75 &  & \nodata  & \nodata  &  \\
{[\ion{Ne}{VI}]} & 7.652 & 126 & 5.80 & 208.4 $\pm$ 1.6 & 127 $\pm$ 15 &  & 12.48 $\pm$ 0.23 & 193 $\pm$ 22 &  & \nodata  & \nodata  &  \\
{[\ion{Fe}{VII}]} & 7.814 & 99.0 & 6.10 & 7.0 $\pm$ 1.6 & \nodata &  & $<$0.39 & \nodata  &  & \nodata  & \nodata  &  \\
{[\ion{Ar}{V}]} & 7.902 & 59.6 & 5.20 & 11.62 $\pm$ 0.85 & \nodata &  & 1.03 $\pm$ 0.23 & \nodata &  & \nodata  & \nodata  &  \\
H$_2$ 0--0 S(4) & 8.026 & \nodata & \nodata & 15.7 $\pm$ 1.0 & 195 $\pm$ 58 &  & 14.44 $\pm$ 0.21 & 178 $\pm$ 26 &  & \nodata  & \nodata  &  \\
{[\ion{Ar}{III}]} & 8.991 & 27.6 & 5.28 & 62.5 $\pm$ 1.1 & 128 $\pm$ 16 &  & 19.02 $\pm$ 0.29 & 264 $\pm$ 12 &  & \nodata  & \nodata  &  \\
{[\ion{Mg}{VII}]} & 9.009 & {187} & 5.87 & $<$2.76 & \nodata  &  & \nodata  & \nodata  &  & \nodata  & \nodata  &  \\
{[\ion{Fe}{VII}]} & 9.527 & 99.0 & 5.74 & 9.73 $\pm$ 0.78 & \nodata &  & 2.27 $\pm$ 0.17 & 298 $\pm$ 77 &  & \nodata  & \nodata  &  \\
H$_2$ 0--0 S(3) & 9.665 & \nodata & \nodata & 39.25 $\pm$ 0.96 & 200 $\pm$ 38 &  & 33.05 $\pm$ 0.29 & 170 $\pm$ 13 &  & \nodata  & \nodata  &  \\
{[\ion{S}{IV}]} & 10.51 & 34.9 & 4.75 & 220.0 $\pm$ 1.9 & 138 $\pm$ 16 &  & 49.94 $\pm$ 0.41 & 260 $\pm$ 6 &  & \nodata  & \nodata  &  \\
\hline
\multicolumn{13}{c}{MIRI\slash MRS Channel 3}  \\
\hline
H$_2$ 0--0 S(2) & 12.28 & \nodata & \nodata & 22.06 $\pm$ 0.92 & 299 $\pm$ 27 &  & 10.51 $\pm$ 0.13 & 141 $\pm$ 16 &  & 1.30 $\pm$ 0.18 & \nodata &  \\
H\,I 7-6 & 12.37 & \nodata & \nodata & $<$1.51 & \nodata  &  & 1.01 $\pm$ 0.10 & 327 $\pm$ 70 &  & $<$0.20 & \nodata  &  \\
{[\ion{Ne}{II}]} & 12.81 & 21.6 & 5.80 & 222.1 $\pm$ 3.3 & 171 $\pm$ 23 &  & 112.8 $\pm$ 1.0 & 311 $\pm$ 6 &  & 8.36 $\pm$ 0.14 & 239 $\pm$ 25 &  \\
{[\ion{Ar}{V}]} & 13.10 & 59.6 & 4.47 & 17.1 $\pm$ 1.5 & \nodata &  & 1.84 $\pm$ 0.21 & \nodata &  & $<$0.23 & \nodata  &  \\
{[\ion{Mg}{V}]} & 13.52 & 109 & 5.70 & $<$2.75 & \nodata  &  & $<$0.22 & \nodata  &  & $<$0.13 & \nodata  &  \\
{[\ion{Ne}{V}]} & 14.32 & 97.2 & 4.51 & 213.4 $\pm$ 1.4 & 126 $\pm$ 9 &  & 24.70 $\pm$ 0.17 & \nodata &  & 5.03 $\pm$ 0.06 & 149 $\pm$ 9 &  \\
{[\ion{Cl}{II}]} & 14.37 & 13.0 & 4.59 & $<$3.66 & \nodata  &  & 1.64 $\pm$ 0.12 & \nodata &  & $<$0.16 & \nodata  &  \\
{[\ion{Ne}{III}]} & 15.56 & 41.0 & 5.32 & 462.5 $\pm$ 3.2 & 154 $\pm$ 12 &  & 131.3 $\pm$ 1.1 & 280 $\pm$ 7 &  & 27.68 $\pm$ 0.33 & 191 $\pm$ 29 &  \\
H$_2$ 0--0 S(1) & 17.03 & \nodata & \nodata & 20.2 $\pm$ 2.4 & \nodata &  & 13.35 $\pm$ 0.66 & \nodata &  & 3.52 $\pm$ 0.21 & \nodata \\
\hline
\multicolumn{13}{c}{MIRI\slash MRS Channel 4}  \\
\hline
{[\ion{Fe}{II}]} & 17.94 & 7.9 & 4.39 & 12.1 $\pm$ 2.1 & 311 $\pm$ 97 &  & 8.35 $\pm$ 0.21 & 366 $\pm$ 15 &  & $<$0.33 & \nodata  &  \\
{[\ion{S}{III}]} & 18.71 & 23.3 & 4.07 & 218.5 $\pm$ 1.8 & 141 $\pm$ 18 &  & 60.43 $\pm$ 0.63 & 242 $\pm$ 10 &  & 12.17 $\pm$ 0.25 & 202 $\pm$ 34 &  \\
{[\ion{Ne}{V}]} & 24.32 & 97.2 & 3.77 & 312.3 $\pm$ 10.4 & 178 $\pm$ 47 &  & 23.81 $\pm$ 0.34 & 194 $\pm$ 29 &  & 5.27 $\pm$ 0.41 & \nodata &  \\
{[\ion{O}{IV}]} & 25.89 & 54.9 & 4.00 & 1167.2 $\pm$ 4.9 & \nodata &  & 127.1 $\pm$ 1.1 & \nodata &  & 48.45 $\pm$ 0.92 & 191 $\pm$ 8 &  \\
{[\ion{Fe}{II}]} & 25.99 & 7.9 & 3.95 & 42.4 $\pm$ 4.3 & \nodata &  & 11.93 $\pm$ 0.67 & \nodata &  & $<$1.90 & \nodata  &  \\

\hline
\end{tabular}
\end{tiny}
\tablefoot{Flux (zeroth moment) and velocity dispersion (second moment; $\sigma$) measured at the AGN, N2, and S2 regions.
The diameter of the apertures is 1\arcsec. A point-source aperture correction is applied to the AGN spectra. The quoted uncertainties are 1$\sigma$ statistical uncertainties. The absolute calibration uncertainty is $\sim$10\%\ \citep{Rigby2022}. For the non-detections, we list the 3$\sigma$ upper limits for a line with a width of $\sim$200\,km\,s$^{-1}$. 
\tablefoottext{a}{Ionization potential of the transition defined as the energy needed to reach the ionization stage producing that transition \citep{NIST_ASD}.}
\tablefoottext{b}{Critical density at $T=10000$\,K for collisions with e$^-$ calculated using PyNeb v1.1.15 \citep{Luridiana2015}.}
}
\end{table*}

\begin{figure*}
\centering
\vspace{5mm}
\includegraphics[width=0.85\textwidth]{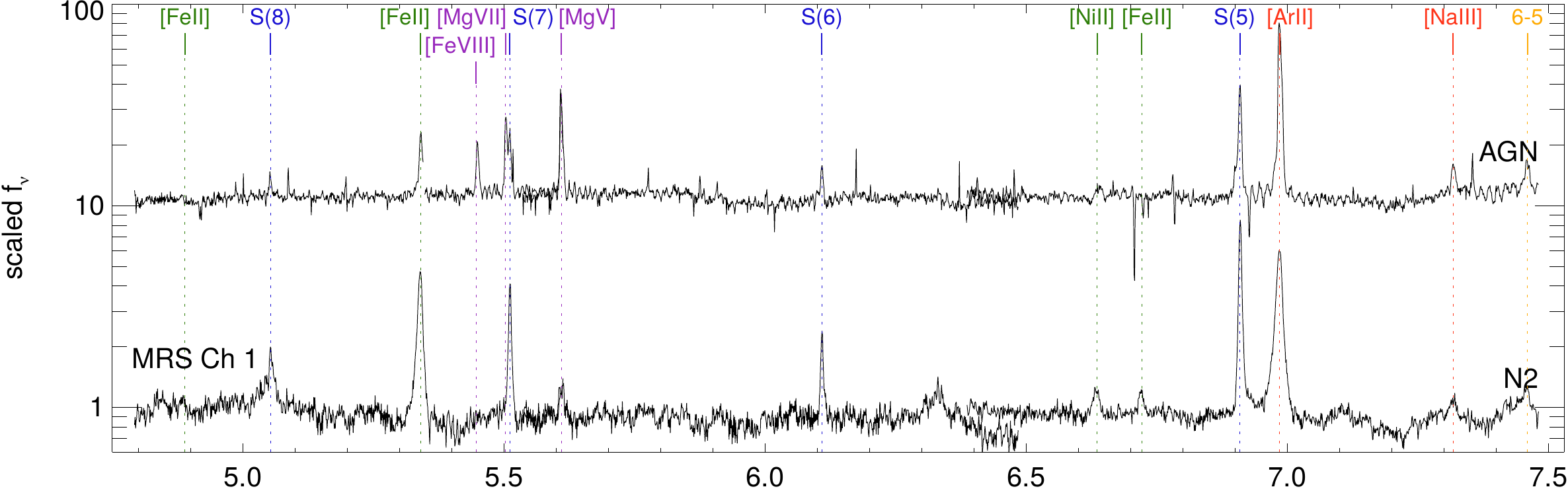}
\includegraphics[width=0.85\textwidth]{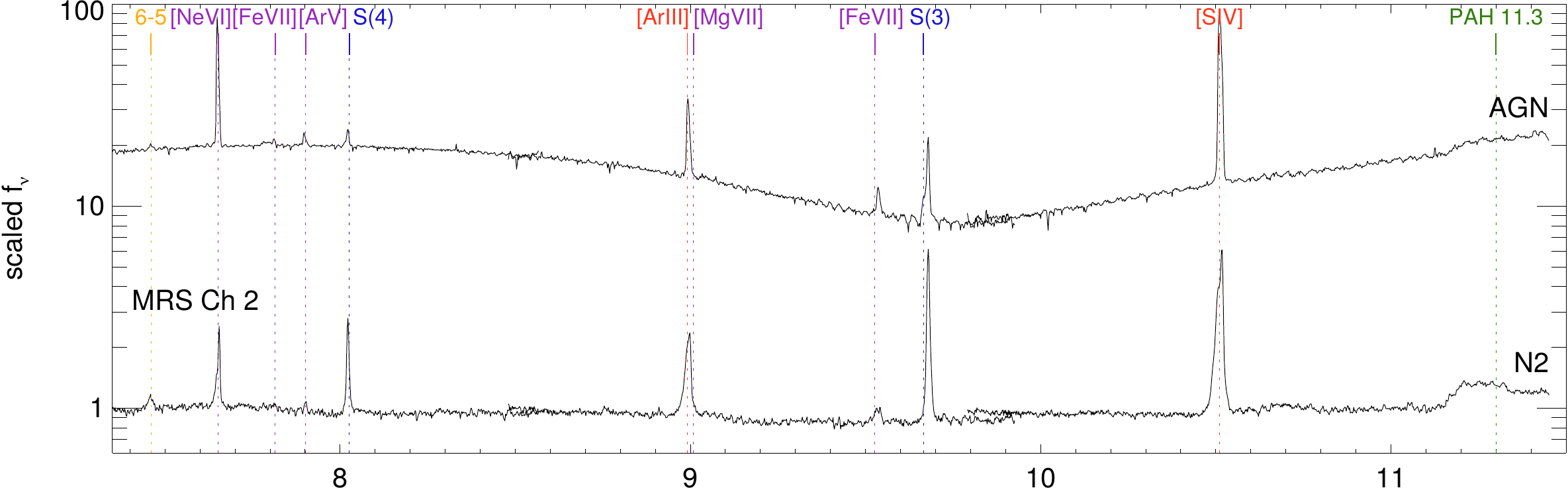}
\includegraphics[width=0.85\textwidth]{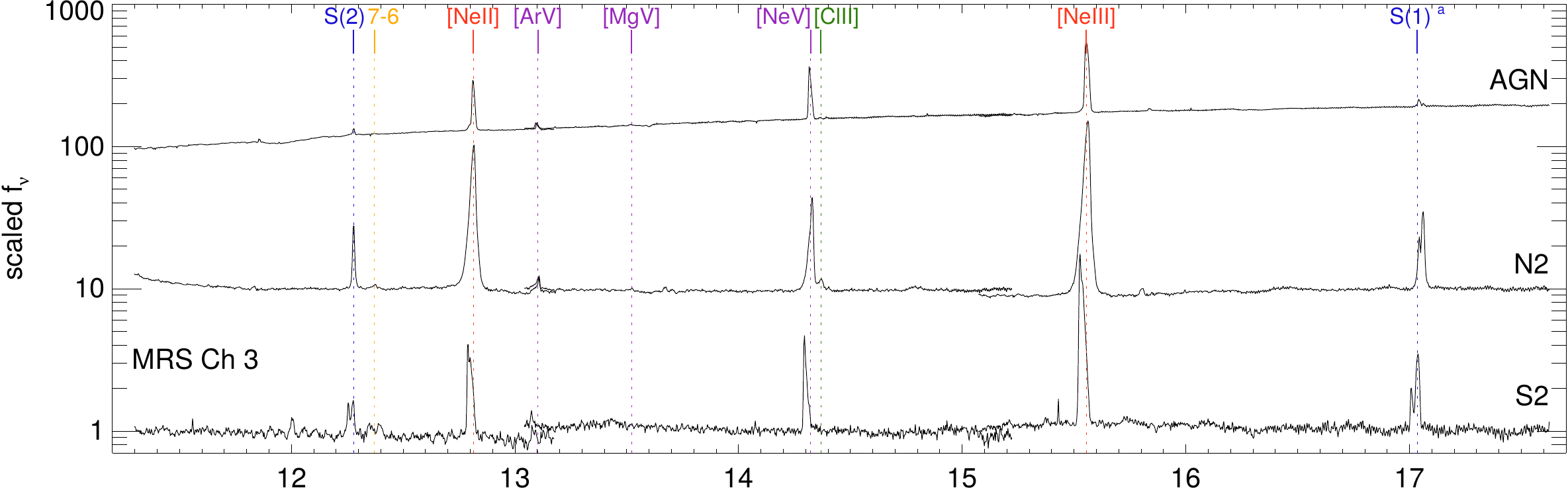}
\includegraphics[width=0.85\textwidth]{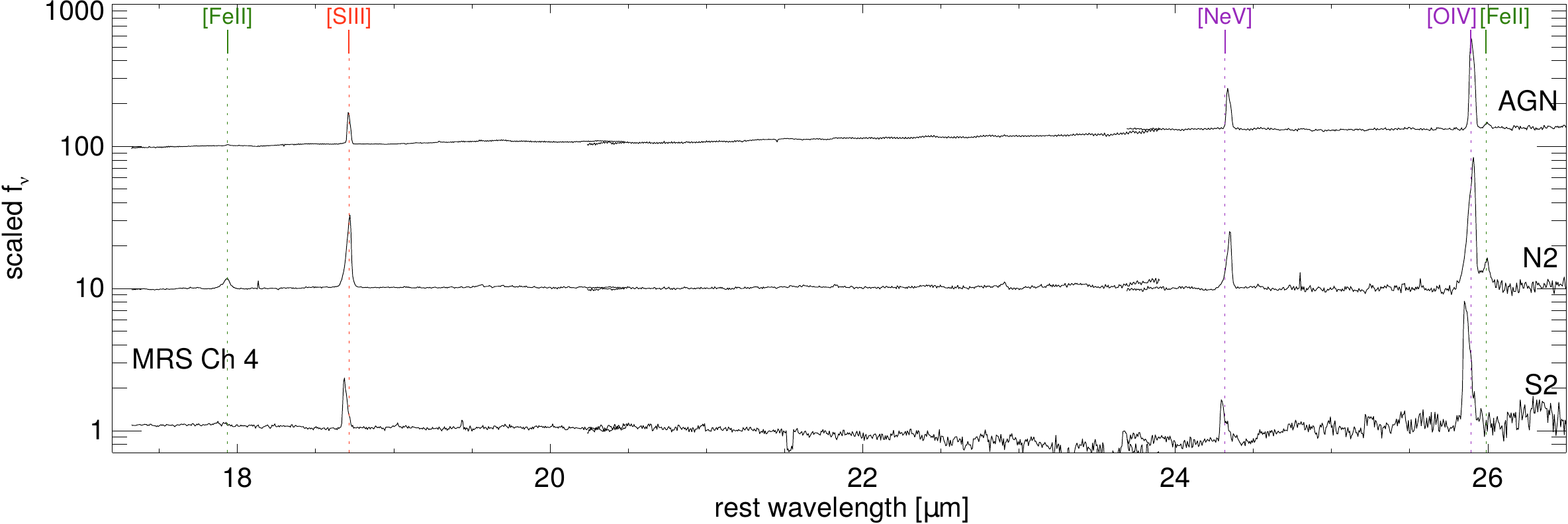}
\caption{Continuum-subtracted MIRI\slash MRS spectra of three regions (AGN, N2, and S2) of NGC~7319 (see Figs.~\ref{fig_mrs_system}, \ref{fig_mrs_N2}, and \ref{fig_mrs_S2}) shown at rest wavelength (assuming $z$ = 0.02251). The continuum level is determined from a linear fit to the flux at the edges of the covered spectral range in each channel.
A point-source aperture correction is applied to the AGN spectra. From top to bottom, the panels show the spectra from MRS channels 1, 2, 3, and 4, respectively. 
Region S2 is not covered by the FoVs of channels 1 or 2. The wavelength of the emission lines are indicated in blue for rotational H$_2$ transitions, in orange for H recombination lines, and in green, red, and purple for transitions from low-ionization (IP$<$13.6\,eV), intermediate-ionization (13.6\,eV$<$IP$<$54.4\,eV), and high-ionization (IP$>$54.4\,eV) species, respectively. The continuum-subtracted spectra have been scaled and shifted for visualization purposes. In this figure we shifted the channel 3 medium sub-band wavelength by 0.04\micron\ to correct for a wavelength calibration artifact (see Appendix~\ref{apx_reduction}).\\
$^a$ The H$_2$ S(1) 17.03\micron\ transition is affected by a wavelength calibration artifact that alters the line profile of the AGN and N2 spectra but not the observed line flux (see Appendix~\ref{apx_reduction}).
\label{fig_mrs_spec}}
\end{figure*}

\onecolumn
\section{Ionized and warm molecular gas velocity field}

\begin{figure}[h]
\centering
\vspace{5mm}
\includegraphics[width=0.6\textwidth]{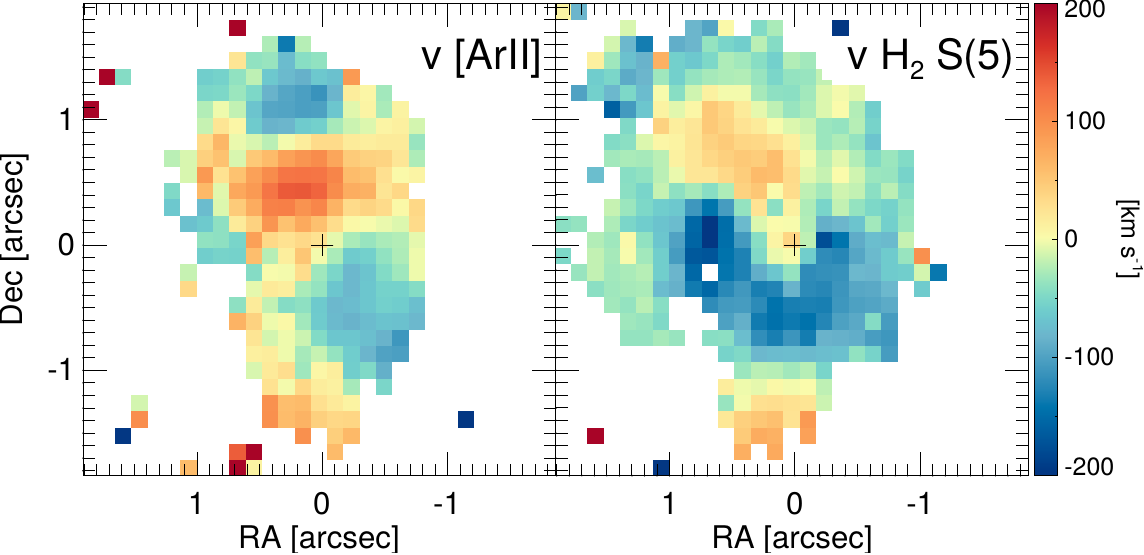}
\caption{Velocity field, first moment, derived from the \Arii\ (left) and H$_2$ S(5) 6.91\micron\ (right) emission lines. The FoV is the same as in Figs.~\ref{fig_mrs_N2} and \ref{fig_mrs_N2_sigma}.
\label{fig_mrs_N2_v}}
\end{figure}

\section{Mid-IR [\ion{Fe}{ii}] density diagnostics}

\begin{figure}[h]
\centering
\vspace{5mm}
\includegraphics[width=0.43\textwidth]{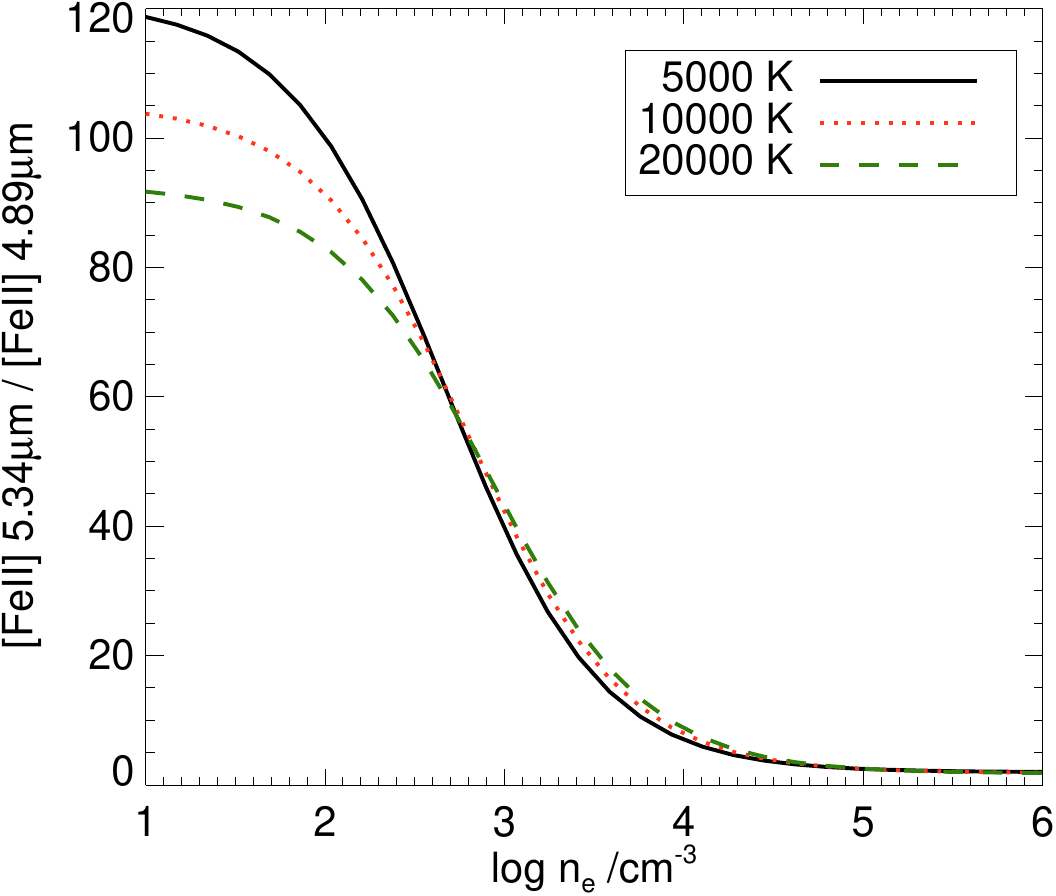}
\caption{\Feii\slash [\ion{Fe}{ii}]4.89\micron\ ratio as a function of the electron density at 5000\,K (solid black line), 10000\,K (dotted red line), and 20000\,K (dashed green line). The [\ion{Fe}{ii}]17.94\micron\ \slash [\ion{Fe}{ii}]4.89\micron\ ratio is {fixed and equal} to 10.9 since both transitions have the same upper level (3d$^7$ a$_4$\,F$_{7\slash2}$). Therefore, both transitions can be used interchangeably to determine the electron density together with the \Feii\ line. The line emissivities are calculated using PyNeb \citep{Luridiana2015} and the Fe$^+$ atomic parameters from \citet{Smyth2019}.
\label{fig_fe2_density}}
\end{figure}


\begin{thebibliography}{51}
\expandafter\ifx\csname natexlab\endcsname\relax\def\natexlab#1{#1}\fi

\bibitem[{{Alatalo} {et~al.}(2011){Alatalo}, {Blitz}, {Young}, {Davis},
  {Bureau}, {Lopez}, {Cappellari}, {Scott}, {Shapiro}, {Crocker},
  {Mart{\'\i}n}, {Bois}, {Bournaud}, {Davies}, {de Zeeuw}, {Duc}, {Emsellem},
  {Falc{\'o}n-Barroso}, {Khochfar}, {Krajnovi{\'c}}, {Kuntschner}, {Lablanche},
  {McDermid}, {Morganti}, {Naab}, {Oosterloo}, {Sarzi}, {Serra}, \&
  {Weijmans}}]{Alatalo2011}
{Alatalo}, K., {Blitz}, L., {Young}, L.~M., {et~al.} 2011, \apj, 735, 88

\bibitem[{{Allen} {et~al.}(2008){Allen}, {Groves}, {Dopita}, {Sutherland}, \&
  {Kewley}}]{Allen2008}
{Allen}, M.~G., {Groves}, B.~A., {Dopita}, M.~A., {Sutherland}, R.~S., \&
  {Kewley}, L.~J. 2008, \apjs, 178, 20

\bibitem[{{Alonso-Herrero} {et~al.}(2018){Alonso-Herrero}, {Pereira-Santaella},
  {Garc{\'\i}a-Burillo}, {Davies}, {Combes}, {Asmus}, {Bunker},
  {D{\'\i}az-Santos}, {Gandhi}, {Gonz{\'a}lez-Mart{\'\i}n},
  {Hern{\'a}n-Caballero}, {Hicks}, {H{\"o}nig}, {Labiano}, {Levenson},
  {Packham}, {Ramos Almeida}, {Ricci}, {Rigopoulou}, {Rosario}, {Sani}, \&
  {Ward}}]{AlonsoHerrero2018}
{Alonso-Herrero}, A., {Pereira-Santaella}, M., {Garc{\'\i}a-Burillo}, S.,
  {et~al.} 2018, \apj, 859, 144

\bibitem[{{{\'A}lvarez-M{\'a}rquez} {et~al.}(2022){{\'A}lvarez-M{\'a}rquez},
  {Labiano}, {Guillard}, {Dicken}, {Argyriou}, {Patapis}, {Law}, {Kavanagh},
  {Larson}, {Gasman}, {Mueller}, {Alberts}, {Brandl}, {Colina},
  {Garc{\'\i}a-Mar{\'\i}n}, {Jones}, {Noriega-Crespo}, {Shivaei}, {Temim}, \&
  {Wright}}]{AlvarezMarquez2022}
{{\'A}lvarez-M{\'a}rquez}, J., {Labiano}, A., {Guillard}, P., {et~al.} 2022,
  arXiv e-prints, arXiv:2209.01695

\bibitem[{{Aoki} {et~al.}(1999){Aoki}, {Kosugi}, {Wilson}, \&
  {Yoshida}}]{Aoki1999}
{Aoki}, K., {Kosugi}, G., {Wilson}, A.~S., \& {Yoshida}, M. 1999, \apj, 521,
  565

\bibitem[{{Aoki} {et~al.}(1996){Aoki}, {Ohtani}, {Yoshida}, \&
  {Kosugi}}]{Aoki1996}
{Aoki}, K., {Ohtani}, H., {Yoshida}, M., \& {Kosugi}, G. 1996, \aj, 111, 140

\bibitem[{{Argyriou} {et~al.}(2020){Argyriou}, {Wells}, {Glasse}, {Lee},
  {Royer}, {Vandenbussche}, {Malumuth}, {Glauser}, {Kavanagh}, {Labiano},
  {Lahuis}, {Mueller}, \& {Patapis}}]{Argyriou2020}
{Argyriou}, I., {Wells}, M., {Glasse}, A., {et~al.} 2020, \aap, 641, A150

\bibitem[{{Beir{\~a}o} {et~al.}(2015){Beir{\~a}o}, {Armus}, {Lehnert},
  {Guillard}, {Heckman}, {Draine}, {Hollenbach}, {Walter}, {Sheth}, {Smith},
  {Shopbell}, {Boulanger}, {Surace}, {Hoopes}, \& {Engelbracht}}]{Beirao2015}
{Beir{\~a}o}, P., {Armus}, L., {Lehnert}, M.~D., {et~al.} 2015, \mnras, 451,
  2640

\bibitem[{{B{\^\i}rzan} {et~al.}(2008){B{\^\i}rzan}, {McNamara}, {Nulsen},
  {Carilli}, \& {Wise}}]{Birzan2008}
{B{\^\i}rzan}, L., {McNamara}, B.~R., {Nulsen}, P.~E.~J., {Carilli}, C.~L., \&
  {Wise}, M.~W. 2008, \apj, 686, 859

\bibitem[{{Boettcher} {et~al.}(2012){Boettcher}, {Harris}, \&
  {Krawczynski}}]{Boettcher2012}
{Boettcher}, M., {Harris}, D.~E., \& {Krawczynski}, H. 2012, {Relativistic Jets
  from Active Galactic Nuclei}

\bibitem[{{Burton} {et~al.}(1992){Burton}, {Hollenbach}, \&
  {Tielens}}]{Burton1992}
{Burton}, M.~G., {Hollenbach}, D.~J., \& {Tielens}, A.~G.~G. 1992, \apj, 399,
  563

\bibitem[{{Dasyra} {et~al.}(2014){Dasyra}, {Combes}, {Novak}, {Bremer},
  {Spinoglio}, {Pereira Santaella}, {Salom{\'e}}, \& {Falgarone}}]{Dasyra2014}
{Dasyra}, K.~M., {Combes}, F., {Novak}, G.~S., {et~al.} 2014, \aap, 565, A46

\bibitem[{{Dav{\'e}} {et~al.}(2019){Dav{\'e}}, {Angl{\'e}s-Alc{\'a}zar},
  {Narayanan}, {Li}, {Rafieferantsoa}, \& {Appleby}}]{Dave2019}
{Dav{\'e}}, R., {Angl{\'e}s-Alc{\'a}zar}, D., {Narayanan}, D., {et~al.} 2019,
  \mnras, 486, 2827

\bibitem[{{Fern{\'a}ndez-Ontiveros} {et~al.}(2020){Fern{\'a}ndez-Ontiveros},
  {Dasyra}, {Hatziminaoglou}, {Malkan}, {Pereira-Santaella}, {Papachristou},
  {Spinoglio}, {Combes}, {Aalto}, {Nagar}, {Imanishi}, {Andreani}, {Ricci}, \&
  {Slater}}]{FernandezOntiveros2020}
{Fern{\'a}ndez-Ontiveros}, J.~A., {Dasyra}, K.~M., {Hatziminaoglou}, E.,
  {et~al.} 2020, \aap, 633, A127

\bibitem[{{Gao} \& {Xu}(2000)}]{Gao2000}
{Gao}, Y. \& {Xu}, C. 2000, \apjl, 542, L83

\bibitem[{{Garc{\'\i}a-Bernete} {et~al.}(2021){Garc{\'\i}a-Bernete},
  {Alonso-Herrero}, {Garc{\'\i}a-Burillo}, {Pereira-Santaella},
  {Garc{\'\i}a-Lorenzo}, {Carrera}, {Rigopoulou}, {Ramos Almeida}, {Villar
  Mart{\'\i}n}, {Gonz{\'a}lez-Mart{\'\i}n}, {Hicks}, {Labiano}, {Ricci}, \&
  {Mateos}}]{GarciaBernete2021}
{Garc{\'\i}a-Bernete}, I., {Alonso-Herrero}, A., {Garc{\'\i}a-Burillo}, S.,
  {et~al.} 2021, \aap, 645, A21

\bibitem[{{Garc{\'\i}a-Bernete} {et~al.}(2022){Garc{\'\i}a-Bernete},
  {Rigopoulou}, {Alonso-Herrero}, {Pereira-Santaella}, {Roche}, \&
  {Kerkeni}}]{GarciaBernete2022}
{Garc{\'\i}a-Bernete}, I., {Rigopoulou}, D., {Alonso-Herrero}, A., {et~al.}
  2022, \mnras, 509, 4256

\bibitem[{{Garc{\'{\i}}a-Burillo} {et~al.}(2019){Garc{\'{\i}}a-Burillo},
  {Combes}, {Ramos Almeida}, {Usero}, {Alonso-Herrero}, {Hunt}, {Rouan},
  {Aalto}, {Querejeta}, {Viti}, {van der Werf}, {Vives-Arias}, {Fuente},
  {Colina}, {Mart{\'\i}n-Pintado}, {Henkel}, {Mart{\'\i}n}, {Krips},
  {Gratadour}, {Neri}, \& {Tacconi}}]{GarciaBurillo2019}
{Garc{\'{\i}}a-Burillo}, S., {Combes}, F., {Ramos Almeida}, C., {et~al.} 2019,
  \aap, 632, A61

\bibitem[{{Garc{\'{\i}}a-Burillo} {et~al.}(2014){Garc{\'{\i}}a-Burillo},
  {Combes}, {Usero}, {Aalto}, {Krips}, {Viti}, {Alonso-Herrero}, {Hunt},
  {Schinnerer}, {Baker}, {Boone}, {Casasola}, {Colina}, {Costagliola},
  {Eckart}, {Fuente}, {Henkel}, {Labiano}, {Mart{\'{\i}}n}, {M{\'a}rquez},
  {Muller}, {Planesas}, {Ramos Almeida}, {Spaans}, {Tacconi}, \& {van der
  Werf}}]{GarciaBurillo2014}
{Garc{\'{\i}}a-Burillo}, S., {Combes}, F., {Usero}, A., {et~al.} 2014, \aap,
  567, A125

\bibitem[{{Guillard} {et~al.}(2012){Guillard}, {Ogle}, {Emonts}, {Appleton},
  {Morganti}, {Tadhunter}, {Oosterloo}, {Evans}, \& {Evans}}]{Guillard2012}
{Guillard}, P., {Ogle}, P.~M., {Emonts}, B.~H.~C., {et~al.} 2012, \apj, 747, 95

\bibitem[{{Hollenbach} \& {McKee}(1989)}]{Hollenbach1989}
{Hollenbach}, D. \& {McKee}, C.~F. 1989, \apj, 342, 306

\bibitem[{{Koo} {et~al.}(2016){Koo}, {Raymond}, \& {Kim}}]{Koo2016}
{Koo}, B.-C., {Raymond}, J.~C., \& {Kim}, H.-J. 2016, Journal of Korean
  Astronomical Society, 49, 109

\bibitem[{Kramida {et~al.}(2021)Kramida, {Yu.~Ralchenko}, Reader, \& {and NIST
  ASD Team}}]{NIST_ASD}
Kramida, A., {Yu.~Ralchenko}, Reader, J., \& {and NIST ASD Team}. 2021, {NIST
  Atomic Spectra Database (ver. 5.9), [Online]. Available:
  {\tt{https://physics.nist.gov/asd}} [2022, July 27]. National Institute of
  Standards and Technology, Gaithersburg, MD.}

\bibitem[{{Labiano} {et~al.}(2021){Labiano}, {Argyriou},
  {{\'A}lvarez-M{\'a}rquez}, {Glasse}, {Glauser}, {Patapis}, {Law}, {Brandl},
  {Justtanont}, {Lahuis}, {Mart{\'\i}nez-Galarza}, {Mueller}, {Noriega-Crespo},
  {Royer}, {Shaughnessy}, \& {Vandenbussche}}]{Labiano2021}
{Labiano}, A., {Argyriou}, I., {{\'A}lvarez-M{\'a}rquez}, J., {et~al.} 2021,
  \aap, 656, A57

\bibitem[{{Le Bourlot} {et~al.}(1999){Le Bourlot}, {Pineau des For{\^e}ts}, \&
  {Flower}}]{LeBourlot1999}
{Le Bourlot}, J., {Pineau des For{\^e}ts}, G., \& {Flower}, D.~R. 1999, \mnras,
  305, 802

\bibitem[{{Luridiana} {et~al.}(2015){Luridiana}, {Morisset}, \&
  {Shaw}}]{Luridiana2015}
{Luridiana}, V., {Morisset}, C., \& {Shaw}, R.~A. 2015, \aap, 573, A42

\bibitem[{{Morganti} {et~al.}(2015){Morganti}, {Oosterloo}, {Oonk},
  {Frieswijk}, \& {Tadhunter}}]{Morganti2015}
{Morganti}, R., {Oosterloo}, T., {Oonk}, J.~B.~R., {Frieswijk}, W., \&
  {Tadhunter}, C. 2015, \aap, 580, A1

\bibitem[{{Mukherjee} {et~al.}(2016){Mukherjee}, {Bicknell}, {Sutherland}, \&
  {Wagner}}]{Mukherjee2016}
{Mukherjee}, D., {Bicknell}, G.~V., {Sutherland}, R., \& {Wagner}, A. 2016,
  \mnras, 461, 967

\bibitem[{{Ogle} {et~al.}(2010){Ogle}, {Boulanger}, {Guillard}, {Evans},
  {Antonucci}, {Appleton}, {Nesvadba}, \& {Leipski}}]{Ogle2010}
{Ogle}, P., {Boulanger}, F., {Guillard}, P., {et~al.} 2010, \apj, 724, 1193

\bibitem[{{Pereira-Santaella} {et~al.}(2010{\natexlab{a}}){Pereira-Santaella},
  {Alonso-Herrero}, {Rieke}, {Colina}, {D{\'{\i}}az-Santos}, {Smith},
  {P{\'e}rez-Gonz{\'a}lez}, \& {Engelbracht}}]{Pereira2010}
{Pereira-Santaella}, M., {Alonso-Herrero}, A., {Rieke}, G.~H., {et~al.}
  2010{\natexlab{a}}, \apjs, 188, 447

\bibitem[{{Pereira-Santaella} {et~al.}(2010{\natexlab{b}}){Pereira-Santaella},
  {Diamond-Stanic}, {Alonso-Herrero}, \& {Rieke}}]{Pereira2010c}
{Pereira-Santaella}, M., {Diamond-Stanic}, A.~M., {Alonso-Herrero}, A., \&
  {Rieke}, G.~H. 2010{\natexlab{b}}, \apj, 725, 2270

\bibitem[{{Pontoppidan} {et~al.}(2022){Pontoppidan}, {Blome}, {Braun}, {Brown},
  {Carruthers}, {Coe}, {DePasquale}, {Espinoza}, {Garcia Marin}, {Gordon},
  {Henry}, {Hustak}, {James}, {Koekemoer}, {LaMassa}, {Law}, {Lockwood},
  {Moro-Martin}, {Mullally}, {Pagan}, {Player}, {Proffitt}, {Pulliam},
  {Ramsay}, {Ravindranath}, {Reid}, {Robberto}, {Sabbi}, \&
  {Ubeda}}]{Pontoppidan2022ERO}
{Pontoppidan}, K., {Blome}, C., {Braun}, H., {et~al.} 2022, arXiv e-prints,
  arXiv:2207.13067

\bibitem[{{Ramos Almeida} {et~al.}(2022){Ramos Almeida}, {Bischetti},
  {Garc{\'\i}a-Burillo}, {Alonso-Herrero}, {Audibert}, {Cicone}, {Feruglio},
  {Tadhunter}, {Pierce}, {Pereira-Santaella}, \& {Bessiere}}]{RamosAlmeida2022}
{Ramos Almeida}, C., {Bischetti}, M., {Garc{\'\i}a-Burillo}, S., {et~al.} 2022,
  \aap, 658, A155

\bibitem[{{Renaud} {et~al.}(2010){Renaud}, {Appleton}, \& {Xu}}]{Renaud2010}
{Renaud}, F., {Appleton}, P.~N., \& {Xu}, C.~K. 2010, \apj, 724, 80

\bibitem[{{Ricci} {et~al.}(2017){Ricci}, {Trakhtenbrot}, {Koss}, {Ueda}, {Del
  Vecchio}, {Treister}, {Schawinski}, {Paltani}, {Oh}, {Lamperti}, {Berney},
  {Gandhi}, {Ichikawa}, {Bauer}, {Ho}, {Asmus}, {Beckmann}, {Soldi},
  {Balokovi{\'c}}, {Gehrels}, \& {Markwardt}}]{Ricci2017_BAT}
{Ricci}, C., {Trakhtenbrot}, B., {Koss}, M.~J., {et~al.} 2017, \apjs, 233, 17

\bibitem[{{Rieke} {et~al.}(2015){Rieke}, {Ressler}, {Morrison}, {Bergeron},
  {Bouchet}, {Garc{\'\i}a-Mar{\'\i}n}, {Greene}, {Regan}, {Sukhatme}, \&
  {Walker}}]{Rieke2015}
{Rieke}, G.~H., {Ressler}, M.~E., {Morrison}, J.~E., {et~al.} 2015, \pasp, 127,
  665

\bibitem[{{Rigby} {et~al.}(2022){Rigby}, {Perrin}, {McElwain}, {Kimble},
  {Friedman}, {Lallo}, {Doyon}, \& {Feinberg}}]{Rigby2022}
{Rigby}, J., {Perrin}, M., {McElwain}, M., {et~al.} 2022, arXiv e-prints,
  arXiv:2207.05632

\bibitem[{{Rodr{\'\i}guez-Baras} {et~al.}(2014){Rodr{\'\i}guez-Baras},
  {Rosales-Ortega}, {D{\'\i}az}, {S{\'a}nchez}, \&
  {Pasquali}}]{RodriguezBaras14}
{Rodr{\'\i}guez-Baras}, M., {Rosales-Ortega}, F.~F., {D{\'\i}az}, A.~I.,
  {S{\'a}nchez}, S.~F., \& {Pasquali}, A. 2014, \mnras, 442, 495

\bibitem[{{Smith} {et~al.}(2007){Smith}, {Draine}, {Dale}, {Moustakas},
  {Kennicutt}, {Helou}, {Armus}, {Roussel}, {Sheth}, {Bendo}, {Buckalew},
  {Calzetti}, {Engelbracht}, {Gordon}, {Hollenbach}, {Li}, {Malhotra},
  {Murphy}, \& {Walter}}]{Smith07}
{Smith}, J.~D.~T., {Draine}, B.~T., {Dale}, D.~A., {et~al.} 2007, \apj, 656,
  770

\bibitem[{{Smyth} {et~al.}(2019){Smyth}, {Ramsbottom}, {Keenan}, {Ferland}, \&
  {Ballance}}]{Smyth2019}
{Smyth}, R.~T., {Ramsbottom}, C.~A., {Keenan}, F.~P., {Ferland}, G.~J., \&
  {Ballance}, C.~P. 2019, \mnras, 483, 654

\bibitem[{{Storey} \& {Hummer}(1995)}]{Storey1995}
{Storey}, P.~J. \& {Hummer}, D.~G. 1995, \mnras, 272, 41

\bibitem[{{Sulentic} {et~al.}(2001){Sulentic}, {Rosado}, {Dultzin-Hacyan},
  {Verdes-Montenegro}, {Trinchieri}, {Xu}, \& {Pietsch}}]{Sulentic2001}
{Sulentic}, J.~W., {Rosado}, M., {Dultzin-Hacyan}, D., {et~al.} 2001, \aj, 122,
  2993

\bibitem[{{Sutherland} \& {Bicknell}(2007)}]{Sutherland2007}
{Sutherland}, R.~S. \& {Bicknell}, G.~V. 2007, \apjs, 173, 37

\bibitem[{{Venturi} {et~al.}(2021){Venturi}, {Cresci}, {Marconi}, {Mingozzi},
  {Nardini}, {Carniani}, {Mannucci}, {Marasco}, {Maiolino}, {Perna},
  {Treister}, {Bland-Hawthorn}, \& {Gallimore}}]{Venturi2021}
{Venturi}, G., {Cresci}, G., {Marconi}, A., {et~al.} 2021, \aap, 648, A17

\bibitem[{{Weinberger} {et~al.}(2017){Weinberger}, {Springel}, {Hernquist},
  {Pillepich}, {Marinacci}, {Pakmor}, {Nelson}, {Genel}, {Vogelsberger},
  {Naiman}, \& {Torrey}}]{Weinberger2017}
{Weinberger}, R., {Springel}, V., {Hernquist}, L., {et~al.} 2017, \mnras, 465,
  3291

\bibitem[{Wells {et~al.}(2015)Wells, Pel, Glasse, Wright, Aitink-Kroes,
  Azzollini, Beard, Brandl, Gallie, Geers, \& et~al.}]{Wells2015}
Wells, M., Pel, J.-W., Glasse, A., {et~al.} 2015, PASP, 127, 646–664

\bibitem[{{Williams} {et~al.}(2002){Williams}, {Yun}, \&
  {Verdes-Montenegro}}]{Williams2002}
{Williams}, B.~A., {Yun}, M.~S., \& {Verdes-Montenegro}, L. 2002, \aj, 123,
  2417

\bibitem[{{Wright} {et~al.}(2015){Wright}, {Wright}, {Goodson}, {Rieke},
  {Aitink-Kroes}, {Amiaux}, {Aricha-Yanguas}, \& {Azzollini}}]{Wright2015}
{Wright}, G.~S., {Wright}, D., {Goodson}, G.~B., {et~al.} 2015, \pasp, 127, 595

\bibitem[{{Xanthopoulos} {et~al.}(2004){Xanthopoulos}, {Muxlow}, {Thomasson},
  \& {Garrington}}]{Xanthopoulos2004}
{Xanthopoulos}, E., {Muxlow}, T.~W.~B., {Thomasson}, P., \& {Garrington}, S.~T.
  2004, \mnras, 353, 1117

\bibitem[{{Yttergren} {et~al.}(2021){Yttergren}, {Misquitta},
  {S{\'a}nchez-Monge}, {Valencia-S}, {Eckart}, {Zensus}, \&
  {Peitl-Thiesen}}]{Yttergren2021}
{Yttergren}, M., {Misquitta}, P., {S{\'a}nchez-Monge}, {\'A}., {et~al.} 2021,
  \aap, 656, A83

\bibitem[{{Zakamska} {et~al.}(2016){Zakamska}, {Lampayan}, {Petric}, {Dicken},
  {Greene}, {Heckman}, {Hickox}, {Ho}, {Krolik}, {Nesvadba}, {Strauss},
  {Geach}, {Oguri}, \& {Strateva}}]{Zakamska2016}
{Zakamska}, N.~L., {Lampayan}, K., {Petric}, A., {et~al.} 2016, \mnras, 455,
  4191

\end{thebibliography}
\end{document}